\documentclass{aa}

\usepackage{graphicx}
\usepackage{cuted}

\usepackage[utf8]{inputenc}

\usepackage{txfonts}
\usepackage[dvipsnames]{xcolor}

\usepackage[colorlinks=true,citecolor=blue]{hyperref}
\usepackage{amsmath}
\usepackage{breqn}
\usepackage{pgfplots}
\usetikzlibrary{quotes,angles}
\usepackage[normalem]{ulem}
\newcommand{\cac}{ c_{\rm Ac}}

\def\nunc{{\nu_{\rm nc 1}}}
\def\nucn{{\nu_{\rm cn 1}}}
\def\psiOne{\psi_{\rm n 1}}
\def\phiOne{\phi_{\rm c 1}}
\def\psiTwo{\psi_{\rm n 2}}
\def\phiTwo{\phi_{\rm c 2}}

\def\gn{g_{\rm n}}
\def\gc{g_{\rm c}}

\def\cAone{c_{\rm Ac1}}

\def \equi#1{\mathrel{\mathop{\kern 0pt\sim}\limits_{#1}}}

\defcitealias{Callies2025}{Paper~I}

\usepackage{subcaption}
\begin{document} 
\makeatletter
\let\linenumbers\relax
\let\endlinenumbers\relax
\makeatother

\title{The Rayleigh–Taylor instability in partially ionized plasmas: ambipolar diffusion effects in the non linear phase}
\titlerunning{Magnetic Rayleigh-Taylor Instability at a Contact Discontinuity}
   
  \author{E. Callies \inst{1},
         Z. Meliani \inst{2},
   			V. Guillet \inst{1}
   		\and	A. Marcowith \inst{1}\
          }
   \institute{Laboratoire Univers et Particules de Montpellier (LUPM) Universit\'e Montpellier, CNRS/IN2P3, CC72, Place Eug\`ene Bataillon, F-34095 Montpellier Cedex5, France
	\and
    Laboratoire d'\'etude de l'Univers et des ph\'enom\`enes eXtr\^emes(LUX), Observatoire de Paris,Universit\'e PSL, Sorbonne Universit\'e, CNRS, 
92190 Meudon, France\\
         \email{edouard.callies@umontpellier.fr} }
\abstract
{The Rayleigh--Taylor (RT) instability is a key mechanism driving mixing and structure formation in stratified astrophysical media. 
In partially ionised environments such as the molecular interstellar medium, including irradiated H$_2$ regions associated with stellar clusters like the Pleiades, ion--neutral coupling and ambipolar diffusion are expected to play a major role in shaping the instability evolution in the presence of gravity and magnetic fields.}
{We aim to determine how ion--neutral coupling and ambipolar diffusion affect the linear and the nonlinear growth of the RT instability under astrophysically relevant conditions, and to identify the coupling regimes in which departures from the classical single--fluid picture become significant.}
{We perform high--resolution two--fluid numerical simulations using the {\tt MPI--AMRVAC} code, spanning a wide range of perturbation wavelengths, coupling strengths, from uncoupled to strongly coupled passing by intermediate or ambipolar diffusion-dominated regimes, and magnetic field configurations. 
The linear theory is revisited using a physically consistent formulation with different ion--neutral coupling strengths across the interface and validated against the simulations. 
We investigate the physics of the instability using morphology--based diagnostics of the mixing layer to compare simulations at equivalent nonlinear stages, complemented by spectral, force, and energy budgets analyses.}
{In the linear regime, theoretical growth rates are recovered over a wide range of wavelengths, from the single--fluid limit to intermediate bi--fluid coupling. In the nonlinear regime, ambipolar diffusion modifies the classical quadratic growth and introduces a coupling--dependent evolution.
For multi--wavelength perturbations, the nonlinear dynamics becomes strongly scale--dependent: intermediate coupling enhances fragmentation in hydrodynamic configurations, while magnetised cases exhibit a non--monotonic reorganisation of the interface, with the smoothest morphologies occurring at intermediate coupling.
Spectral and energetic diagnostics indicate that these behaviours correlate with changes in the relative contributions of ion--neutral drift and magnetic stresses during the nonlinear evolution.}
{Our results demonstrate that ambipolar diffusion does not merely rescale RT growth rates, but reshapes the nonlinear, multi--scale dynamics by altering how gravity--driven kinetic energy is redistributed through magnetic tension and ion--neutral drift. 
In magnetised configurations, the magnetic field suppresses small--scale corrugations, while ambipolar diffusion weakens this constraint by allowing partial decoupling between ions and neutrals.}


\maketitle
\section{Introduction}

The Rayleigh–Taylor instability (RTI) is a fundamental hydrodynamic instability that occurs when a dense fluid is supported against gravity by a lighter one. First identified by Lord Rayleigh \citet{Rayleigh1883} and later confirmed experimentally by Taylor \cite{Taylor1950}, it arises when an interface between two fluids of different densities becomes unstable under gravity, with the heavier fluid lying above the lighter one. Small perturbations of the interface are amplified as dense fluid sinks and light fluid rises, leading in the nonlinear regime to the formation of characteristic spikes and bubbles. In its classical formulation, the RTI is a purely hydrodynamic, large-scale instability, independent at leading order of viscosity and compressibility, and is usually analyzed for two immiscible, incompressible, inviscid fluids separated by a planar interface in a uniform gravitational field.

Beyond its idealized formulation, the Rayleigh–Taylor instability is of broad relevance in astrophysics because it naturally arises in any stratified system subjected to an effective acceleration, whether due to gravity, pressure gradients, or inertial forces. It is therefore expected to operate over a wide variety of macroscopic environments, ranging from accretion flows onto compact, magnetized objects \citep{Arons1976,Wang1983} to buoyant structures in galaxy clusters \citep{Robinson2004,Ruszkowski2007}, as well as in solar and stellar contexts where magnetic flux systems evolve under gravity \citep{Isobe2005}. The instability is also a key ingredient in the dynamics of supernova remnants and pulsar wind nebulae, where accelerated dense shells interact with lighter surrounding media \citep{Jun1996a,Hester1996,Bucciantini2004}. In all these systems, the Rayleigh–Taylor instability governs the growth of large-scale structures, promotes mixing between plasma components, and strongly influences the global morphology and evolution of the flow. This ubiquity highlights the RTI as a fundamental mechanism for momentum and energy redistribution in accelerated astrophysical plasmas, well beyond the idealized hydrodynamic setting.

A large body of analytical work has been devoted to understanding the linear phase of the instability. In hydrodynamics, the classical dispersion relation predicts an unbounded growth rate that increases with the square root of the wavenumber, implying the absence of any short-wavelength cutoff. In contrast, Chandrasekhar’s seminal analysis of the magnetohydrodynamic (MHD) RTI \citet{Chandrasekhar1961} showed that a magnetic field tangential to the interface introduces a stabilizing magnetic tension, leading to the existence of a critical wavenumber above which perturbations are stabilized. Later studies extended this framework by considering inclined magnetic fields. In particular, \citet{Vickers}  demonstrated that when the magnetic field is oblique with respect to the interface, the stabilizing effect is weakened and the cutoff can disappear, restoring unstable modes at all wavelengths. In partially ionized plasmas, a bi-fluid description is required to account for the relative drift between ions and neutrals and for non-ideal effects such as ambipolar diffusion. In this context, \citet{Diaz2012} and \citet{Diaz2014} investigated the impact of compressibility and ion–neutral coupling on the linear RTI, showing that partial ionization can significantly modify growth rates and mode structure compared to single-fluid MHD predictions. In our recent linear analysis \citep{Callies2025} (hereafter \citetalias{Callies2025}), we have combined these ingredients by considering the RTI in a bi-fluid framework with an inclined magnetic field. Our results demonstrate that ambipolar diffusion plays a non-negligible role in the linear development of the instability. In particular, in the parameter regime relevant to partially ionized plasmas, the growth rate exhibits a markedly different scaling, increasing approximately linearly with the wavenumber $k$ , rather than following the standard hydrodynamic $k^{1/2} $ dependence. 

While linear analytical studies have provided invaluable insight into the onset and early evolution of the RTI, they are inherently restricted to the initial exponential growth phase. Extending such approaches into the fully non-linear regime is notoriously difficult, as mode coupling, secondary instabilities, and turbulent mixing rapidly invalidate linear assumptions. For this reason, numerical simulations have become the primary tool to investigate the late-time development and saturation of the RTI. A qualitative criterion for the transition to non-linearity is that the interface displacement in the vertical direction becomes comparable to the inverse wavenumber, $\sim 1/k$, that is, when vertical and horizontal scales are of the same order \citep{Fermi1953}; see also the discussion in \citet{Hillier2016}. In the non-linear regime, the hydrodynamic RTI is observed to evolve in a self-similar manner, with the thickness $h$ of the mixing layer governed by
\begin{equation}
h \simeq \alpha A g t^2,
\end{equation}
where $A$ is the Atwood number, $g$ the acceleration, and $\alpha$ a dimensionless non-linear growth rate that is found to be only weakly sensitive to initial conditions \citep{Ristorcelli2004,Cook2004}. This behaviour has been confirmed in numerous laboratory experiments and numerical studies.

In magnetized plasmas, the non-linear phase has been explored through both two- and three-dimensional MHD simulations. Early 2D studies investigated how magnetic fields aligned either tangentially or normally to the interface modify the growth and morphology of the instability \citep{Jun1995}. Fully 3D simulations later demonstrated that magnetic tension can strongly affect bubble and spike structures and the efficiency of mixing, sometimes even enhancing the late-time growth relative to the purely hydrodynamic case by suppressing secondary Kelvin--Helmholtz roll-ups \citep{Stone2007a,Stone2007b}.  , idealised 3D MHD simulations by \citet{Carlyle2017} systematically examined the effect of magnetic field strength in a parameter regime relevant to astrophysical plasmas, showing that stronger fields tend to reduce the non-linear growth rate of rising bubbles and introduce asymmetries between bubbles and spikes, while the overall evolution remains compatible with a self-similar $h \propto A g t^2$ scaling.

Recent numerical studies have significantly advanced the investigation of the nonlinear RTI in solar prominence conditions. Using 2D two-fluid simulations, \citet{Braileanu2021a} explored the influence of magnetic field strength, shear, and mass loading on the nonlinear evolution of the instability at a smooth prominence–corona interface, showing that magnetic shear can substantially reduce or suppress the instability and that nonlinear development is accompanied by coherent magnetic structuring and partial ion–neutral decoupling. In a follow-up study, \citet{Braileanu2021b} demonstrated that the ion–neutral collision frequency critically controls both linear growth rates and the emergence of small-scale structures in the nonlinear regime.  , \citet{Braileanu2023} emphasized magnetic amplification and current-sheet formation driven by the RTI in partially ionized prominence plasmas, highlighting the feedback between instability-induced flows and magnetic topology evolution. These works provide a detailed description of the RTI in realistic solar prominence environments, including smooth interfaces and complex thermodynamics. In contrast, the present study adopts a deliberately idealized configuration with a sharp interface and controlled parameter variations, designed to isolate the dynamical and energetic role of ambipolar diffusion across distinct ion–neutral coupling regimes and to directly connect linear dispersion properties to nonlinear mixing-layer evolution.

Beyond prominence-specific setups, \citet{Changmai2023} performed high-resolution 2.5D ideal MHD simulations to follow the RTI into a fully developed turbulent state, showing that RTI-driven dynamics can generate anisotropic MHD turbulence with coherent field-aligned flows and power-law spectra. At the same time, recent fully three-dimensional incompressible MHD simulations by \citet{Kalluri2025} demonstrated that the nonlinear magnetic RTI can evolve differently in two and three dimensions, due to the presence in 3D of mixed and interchange modes that enhance small-scale mixing and modify energy partition. These results highlight the sensitivity of nonlinear RTI dynamics to magnetic topology and dimensionality. In this context, controlled two-dimensional studies remain valuable for isolating specific physical mechanisms—particularly when the objective is not to reproduce fully developed turbulence but to quantify how ion–neutral coupling and ambipolar diffusion modify growth laws, force balance, and energy redistribution relative to ideal MHD benchmarks.

More generally, as emphasized in the review by \citet{Soler2022}, most nonlinear studies of partially ionized plasmas rely either on ideal MHD or on single-fluid formulations including ambipolar diffusion, while fully nonlinear two-fluid investigations remain comparatively scarce. In parallel, \citet{Khomenko2025} showed that ambipolar diffusion and ion–neutral drift can strongly influence small-scale structuring and energy dissipation in partially ionized solar plasmas when adequately resolved, underscoring the importance of multi-fluid effects for nonlinear plasma dynamics.

Besides, recent high-resolution observations have started to probe the structure of the cold neutral medium at unprecedented spatial scales. Using JWST/NIRCam data, \citet{Vigoureux2026} reported pronounced anisotropy in the two-dimensional Fourier statistics of dust emission in the Pleiades nebula, down to scales of order $\sim 40$~au. The observed scale-dependent anisotropy, aligned with the magnetic field, provides direct evidence that partially ionized, magnetized plasmas can sustain strongly anisotropic small-scale structuring. Although these observations do not identify a specific driving instability, they highlight the need to understand how magnetically mediated processes and ion–neutral coupling shape nonlinear plasma dynamics across scales. In this context, buoyancy-driven instabilities such as the magnetic Rayleigh–Taylor instability, modified by ambipolar diffusion, offer a controlled framework to investigate mechanisms capable of generating anisotropic structuring and scale-dependent energy redistribution in partially ionized environments.
.

Despite these advances, a systematic and controlled exploration of how ambipolar diffusion reshapes the nonlinear self-similar growth of the magnetic RTI across distinct ion–neutral coupling regimes remains lacking. In particular, whether the classical $h \propto A g t^2$ scaling persists, is delayed, or is fundamentally modified in magnetized, partially ionized plasmas has not been clarified in a bi-fluid framework explicitly designed to connect linear theory and nonlinear evolution. In this work, we address this gap by performing high-resolution 2D two-fluid simulations of the magnetic RTI with an oblique field, combining a self-similar analysis based on the normalized mixing height $h/\lambda$ with controlled perturbations spanning multiple wavelengths and coupling regimes. We introduce a lag-time formalism to enable consistent comparison with single-fluid reference cases and show that ambipolar diffusion does not merely rescale the classical $h\propto t^2$ behaviour, but modifies the nonlinear dynamics in a regime-dependent manner, enhancing small-scale activity in the hydrodynamic limit while suppressing it in magnetized configurations through a redistribution of magnetic tension and drag.

The paper is organized as follows. In Sect.~\ref{sec:Model}, we describe the physical model and numerical setup. Sect.~\ref{sec:linear_verif} details the diagnostics used to define the mixing layer as well as the verification of the theoretical analysis performed in  \citepalias{Callies2025}. In Sect.~\ref{sec:NonLin}, we discuss the dependence of the nonlinear evolution on wavelength and coupling strength. . Finally, Sect.~\ref{sec:Conclusion} summarizes our main conclusions and outlines perspectives for future work.

\section{Method}
\label{sec:Model}
The aim of this work is to quantify how ambipolar diffusion modifies the nonlinear development of the Rayleigh--Taylor (RT) instability in a partially ionized plasma with an oblique magnetic field. Building on the linear analysis of \citetalias{Callies2025}, we extend the study beyond the exponential stage and investigate how ion--neutral drift reshapes the structure, growth, and self--similar evolution of the mixing layer once nonlinear interactions dominate. We solve the two--fluid equations for charged and neutral species in a stratified configuration under gravity, exploring parameter regimes where magnetic tension, buoyancy, and ion--neutral coupling compete. The selected wavelengths and coupling strengths are guided by the linear predictions of \citetalias{Callies2025}. We first validate the numerical implementation against the theoretical growth rates before analysing the nonlinear multi--scale evolution. Our simulations are not designed to reproduce a specific observational configuration, but to isolate the physical mechanisms controlling ion--neutral drift and ambipolar diffusion in nonlinear buoyancy-driven mixing. Resolving these effects numerically requires working in parameter regimes where the relevant coupling and diffusion scales are accessible at finite resolution. Our approach is therefore mechanistic: we identify robust trends in how ambipolar diffusion modifies nonlinear RT evolution, which can then inform astrophysical interpretations.

\subsection{Governing two-fluid equations}
We consider a two-fluid system composed of a charged component ($c$) and a neutral component ($n$).  The magnetic field $\mathbf{B}$ is inclined by an angle $\theta$ with respect to the interface.  Ionisation and recombination, Ohmic resistivity, the Hall effect, and explicit resistive heating are neglected.  The governing equations correspond to the two-fluid formulation implemented in \texttt{MPI-AMRVAC} \citep{Braileanu_Keppens_2022AA...664A..55B}, with the specific choice adopted here that the gravitational acceleration acts only on the charged component.  As a result, the neutral fluid does not experience gravity directly and is accelerated solely through ion--neutral collisional coupling.  This asymmetric forcing is consistent with the configuration analysed in \citetalias{Callies2025}and allows us to isolate the role of ion--neutral momentum exchange in the buoyancy-driven evolution of the instability.  The governing equations read:
\begin{align}
\frac{\partial \rho_n}{\partial t}
+ \nabla \cdot (\rho_n \mathbf{v}_n)
&= 0, \\
\frac{\partial \rho_c}{\partial t}
+ \nabla \cdot (\rho_c \mathbf{v}_c)
&= 0, \\
\frac{\partial (\rho_n \mathbf{v}_n)}{\partial t}
+ \nabla \cdot \left( \rho_n \mathbf{v}_n \mathbf{v}_n + p_n \mathbf{I} \right)
&= \mathbf{R}_n, \\
\frac{\partial (\rho_c \mathbf{v}_c)}{\partial t}
+ \nabla \cdot \left[
\rho_c \mathbf{v}_c \mathbf{v}_c
+ \left(p_c + \frac{1}{2} B^2 \right)\mathbf{I}
- \mathbf{B}\mathbf{B}
\right]
&= \rho_c \mathbf{g}_c - \mathbf{R}_n, \\
\frac{\partial e_n^{\mathrm{tot}}}{\partial t}
+ \nabla \cdot
\big[
\mathbf{v}_n ( e_n^{\mathrm{tot}} + p_n )
\big]
&=  M_n, \\
\frac{\partial e_c^{\mathrm{tot}}}{\partial t}
+ \nabla \cdot \Big[
\mathbf{v}_c ( e_c^{\mathrm{tot}} + p_c + \tfrac{1}{2} B^2 )
- \mathbf{B} (\mathbf{v}_c \cdot \mathbf{B})\Big]
&= \rho_c \mathbf{v}_c \cdot \mathbf{g}_c - M_n, \\
\frac{\partial \mathbf{B}}{\partial t}
+ \nabla \cdot \Big[
\mathbf{v}_c \mathbf{B}
- \mathbf{B}\mathbf{v}_c 
\Big]
&= 0,
\end{align}
with collisional momentum and energy exchange terms
\begin{align}
\mathbf{R}_n &=
\rho_n \rho_c \alpha \left( \mathbf{v}_c - \mathbf{v}_n \right), \\
M_n &=
\frac{1}{2} \left( v_c^2 - v_n^2 \right) \rho_n \rho_c \alpha
+ \frac{1}{\gamma - 1}
\left( T_c - T_n \right) \rho_n \rho_c \alpha .
\end{align}
Here $\alpha$ is the constant collision coefficient, so that the collision frequencies satisfy $\nu_{\rm nc} = \alpha / \rho_c$ and $\nu_{\rm cn} = \alpha / \rho_n$ (for the charges and neutrals respectively). We focus on the asymmetric configuration where only the charged fluid experiences gravity, $\mathbf{g}_c = -g\,\hat{\mathbf{z}}$ and $g_n=0$ as discussed in \citetalias{Callies2025}.

\subsection{Initial equilibrium configuration}
The domain consists of two uniform layers separated by a sharp interface at $z=0$, with identical density contrasts for both fluids:
\begin{align}
\mathcal{A} = \frac{\rho_{c,2}-\rho_{c,1}}{\rho_{c,2}+\rho_{c,1}}
             = \frac{\rho_{n,2}-\rho_{n,1}}{\rho_{n,2}+\rho_{n,1}}
             = \frac{3}{5}.
\end{align}

A constant gravity $\mathbf{g}_c=-g\,\hat{\mathbf{z}}$ is applied to the charged fluid. The pressure of the charged fluid satisfies hydrostatic equilibrium,
\begin{align}
\nabla p_c = \rho_c \mathbf{g}_c,
\end{align}
with $p_c(z_{\rm max}) = 2 n_c k_B T_c$, while the neutral pressure is uniform, $p_n = n_n k_B T_n$. We set $T_c=T_n=100\,{\rm K}$ at the top of the box. The magnetic field strength is fixed from the interface pressure through
\begin{align}
|\mathbf{B}|^2 = \mu \frac{p_c(z=0)}{\beta} \ ,
\end{align}

where $\mu$ is the vacuum permeability and $\beta$ the ratio between thermal and magnetic pressure, i.e the beta plasma. Thereby fixing the magnetic field strength from the interface pressure and the chosen value of $\beta$.

\subsection{Numerical setup}

The simulations are performed with the two-fluid module of the open-source code \texttt{MPI-AMRVAC}\footnote{ MPI-AMRVAC v3.0-289-g034bdf3d (commit \texttt{034bdf3d}), compiled and run with OpenMPI 5.0.8.}, parallelised with MPI. We adopt a 2D  configuration. The computational domain is discretised on a $64\times 64$ base grid and refined dynamically through adaptive mesh refinement (AMR). Depending on the simulation, between five and six AMR levels are used, selected through a Löhner-type refinement criterion based on gradients of key physical variables. This allows the mesh to concentrate around the evolving interface and to resolve the fine-scale structures that develop during the instability growth. With six refinement levels, the effective resolution reaches $4096$ cells per direction.

Time integration is carried out with the three-step IMEX--ARS3 scheme \citep{ASCHER1997}, which treats stiff collisional terms implicitly and ideal MHD fluxes explicitly. Spatial fluxes are computed with the HLLD Riemann solver \citep{TVD_RUUTH,2005JCoPh.208..315M} and third-order PPM reconstruction \citep{COLELLA1984174}. A Courant number of $0.8$ is used for the explicit part of the scheme. The divergence of the magnetic field is controlled through a multigrid cleaning method.

Finally, we impose periodic boundary conditions in the horizontal ($x$) direction. In the vertical direction (perpendicular to the interface), the upper and lower boundaries are treated with fixed-value (non-periodic) conditions for all primitive variables. This choice prevents the appearance of vanishing or numerically unstable pressures at the domain edges, which would otherwise result from a naïve extrapolation of the hydrostatic profiles. The vertical boundaries therefore act as reservoirs that maintain the prescribed density, pressure, and magnetic field values, while still allowing the instability to develop freely within the interior of the domain. 

\subsection{Initial perturbations and physical parameter space}
\label{subsec:initial_conditions_and_parameters}

The Rayleigh--Taylor instability is initiated by imposing a small perturbation of the interface position in the horizontal direction. In continuous form, the displacement of the interface is written as
\begin{align}
\eta(x) \propto \eta_0 \sum_{i=n_{\rm osc}}^{N+n_{\rm osc}} c_{i}\,
\sin\!\left(\frac{2\pi i}{L_x}x + \phi_i\right),
\end{align}
 where $L_x=x_{\rm max}-x_{\rm min}$ denotes the horizontal size of the computational domain,  $\eta_0$ is the amplitude of the perturbation, $N$ is the number of excited modes, and $n_{\rm osc}$ is chosen such that $L_x/n_{\rm osc}$ corresponds to the largest injected wavelength. The coefficients $c_i$ and phases $\phi_i$ are drawn from uniform random distributions, with $c_i\in[-1,1],\phi_i\in[0,2\pi[$, and are read explicitly from the input data so that the perturbation spectrum is fully controlled and reproducible. 

To characterize the simulations and to explore the influence of ambipolar diffusion, magnetic tension, and coupling strength, we adopt a strategy in which a subset of physical parameters is fixed while a controlled set of others is varied systematically. The quantities that remain constant throughout the study are summarized in Table~\ref{tab:Constant_parameters}. They define the global normalization of the problem, including the size of the computational domain, the gravitational acceleration applied to the charged component, and the masses and relative abundances of neutrals and charges. The density ratio $n_c/n_n=10^{-4}$ places the system in a weakly ionized regime representative of cold neutral medium conditions, while allowing ion--neutral drift and ambipolar diffusion to play a dynamically significant role. All spatial scales are expressed relative to the box size $L_x$, so that perturbation wavelengths are naturally given as fractions of $L_x$. In particular, the largest injected wavelength is fixed to $\lambda_{\max}=L_x/4$, which ensures that several unstable modes fit within the computational domain while avoiding direct interaction with the vertical boundaries during the linear stage.

\begin{table}
    \centering
    \begin{tabular}{c|c}
        Name of the quantity & Value   \\
        \hline
        Size of the box $L_x$ & $3\times10^{10}\ \mathrm{cm}$\\
        Gravity $g_c$ & $10^{3}\ {\rm cm\,s}^{-2}$\\
        Mass of neutrals $m_n$ & $m_n=m_p$\\
        Mass of charges $m_c$ & $m_c=12\,m_p$\\
        Ratio of density $n_c/n_n$ & $10^{-4}$\\
    \end{tabular}
    \caption{Constant physical parameters used in all simulations. }
    \label{tab:Constant_parameters}
\end{table}

We stress that the absolute values of $L_x$ and $g_c$ are not intended to reproduce a specific Cold Neutral Medium (CNM) configuration.  The box size $L_x$ is chosen so as to rescale the problem to numerically tractable spatial and temporal scales, allowing the instability to develop within affordable computational times while preserving the relevant dimensionless parameters.  Since the governing equations are invariant under appropriate rescaling, the dynamics primarily depends on ratios of characteristic length and time scales rather than on their absolute values. In practice, our strategy is to first fix the magnetic field strength, expressed through the plasma beta parameter $\beta$, which controls the importance of magnetic tension relative to thermal pressure. Unless otherwise stated, we adopt a reference value $\beta = 3 \times 10^{2}$, corresponding to a moderately magnetized regime in which magnetic tension is dynamically important but does not fully suppress the Rayleigh--Taylor instability. Once $\beta$ is fixed, the gravitational acceleration $g_c$ is treated as a free control parameter and is adjusted so as to place the system in a regime where magnetic effects selectively stabilize small--scale modes.  More precisely, $g_c$ is chosen such that the magnetic cutoff wavelength predicted by linear theory satisfies $L_{\rm cut} \simeq \frac{L_x}{20},$ensuring that both magnetically stabilized and unstable modes coexist within the computational domain. This choice maximizes the sensitivity of the nonlinear evolution to ambipolar diffusion effects by enforcing a competition between buoyancy, magnetic tension, and ion--neutral coupling on comparable spatial scales. The resulting parameter regime is therefore deliberately designed to probe how ambipolar diffusion reshapes the nonlinear cascade, rather than to reproduce a specific astrophysical object.
\begin{table}
    \centering
    \begin{tabular}{c|c}
        Name of the quantity & Value   \\
        \hline
        Magnetic $\beta$ & $3\times10^2$\\
        Atwood number $\mathcal{A}$ & $3/5$\\
        Magnetic inclination in the $xOz$ plane $\theta$ & $10^\circ$\\
        Largest injected wavelength $\lambda_{\max}$ & $L_x/4$\\
    \end{tabular}
    \caption{Default values of the parameters varied in the simulations.}
    \label{tab:variable_parameters}
\end{table}

\begin{figure}
    \centering
    \includegraphics[width=\linewidth]{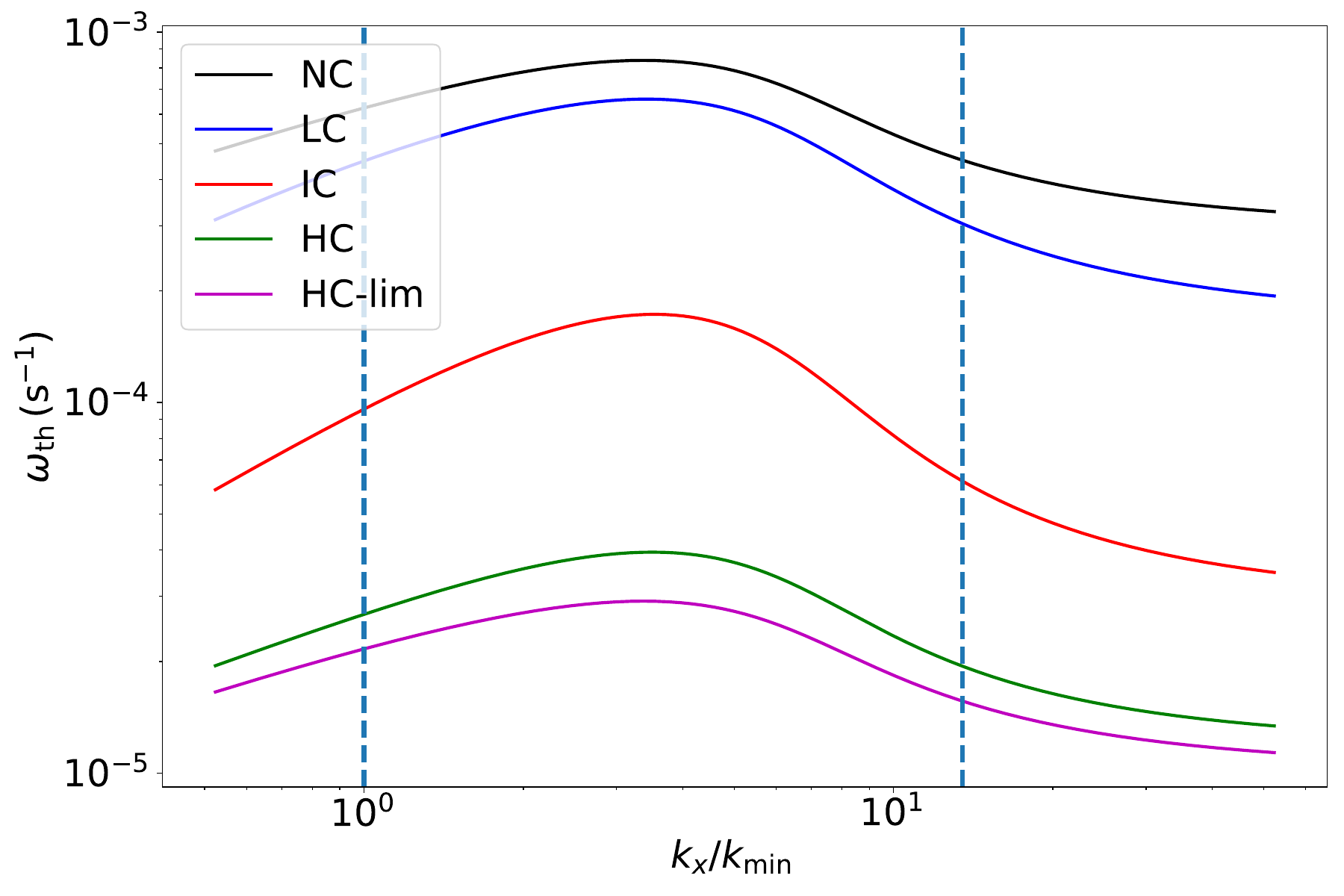}
    \caption{
Theoretical linear growth rate as a function of the horizontal wavenumber in the presence of a magnetic field ($\theta=10^\circ$). Black curves indicate the asymptotic limits corresponding to the fully uncoupled and fully coupled regimes. Colored curves show intermediate coupling cases for two different assumptions on the variation of the collision frequency across the interface. Dashed bands mark the ranges of wavelengths injected in the simulations, illustrating how the numerical experiments sample the different coupling regimes predicted by linear theory.}
    \label{fig:theoretical}
\end{figure}

The role of ion--neutral coupling is characterized through the neutral--ion collision frequency, which is treated as a free parameter and varied over several orders of magnitude. For clarity, coupling regimes are classified according to the ratio $\nu_{\rm nc}/\omega_{\rm th}$, where $\omega_{\rm th}$ (see \citetalias{Callies2025} )  is the theoretical linear growth rate associated with the considered wavelength. This naturally defines four regimes (see Tab. \ref{tab:coupling_regimes} for a summary): no coupling (NC), low coupling (LC), intermediate coupling (IC), high coupling (HC) and high limit coupling (HC-Lim). Figure~\ref{fig:theoretical} illustrates the corresponding theoretical growth rates as functions of the wavenumber for different assumptions regarding the variation of collision frequencies across the interface. In particular, it highlights how adopting a physically consistent variation of $\nu_{\rm nc}$ with density effectively shifts the system toward stronger coupling, without altering the asymptotic limits of the dispersion relation. The  scales between the vertical dashed lines in this figure indicate the ranges of wavelengths injected in the simulations, making explicit how the numerical experiments sample the different coupling regimes predicted by linear theory. The overall survey strategy is therefore twofold: first, to validate the linear behaviour derived in \citetalias{Callies2025} within this physically consistent parameterization; and second, to explore how ambipolar diffusion reshapes the nonlinear evolution of the instability across scales, from single--mode configurations to fully broadband perturbations.

\begin{table*}
    \centering
    \begin{tabular}{c|c|p{10cm}}
        Coupling regime & $\nu_{\rm nc}/\rho_c\ \,[\mathrm{cm^3\,g^{-1}\,s^{-1}}]$ & Physical interpretation \\
        \hline
        NC (No Coupling) &
        $0$ &
        Neutrals and charges evolve independently; no momentum exchange occurs between the two fluids. \\[0.2cm]

        LC (Low Coupling) &
        $7.5\times10^{19}$ &
        Weak coupling regime: ion--neutral interactions are present but insufficient to enforce coherent motion; partial decoupling persists. \\[0.2cm]

        IC (Intermediate Coupling) &
        $7.5\times10^{20}$ &
        Transitional regime: ion--neutral interactions are dynamically important but do not impose a single--fluid behaviour; \bf{ambipolar diffusion effects are maximal}. \\[0.2cm]

        HC (High Coupling) &
        $7.5\times10^{21}$ &
        Strong coupling regime approaching the fully coupled asymptotic limit; charges and neutrals move nearly coherently as a single fluid.\\
        
        HC-Lim (High Coupling Limit) &
        $7.5\times10^{22}$ &
       Totally coupled regime
    \end{tabular}
    \caption{
    Definition of the four ion--neutral coupling regimes used throughout this work.
    The regimes are classified according to the effective collision frequency per unit charged density, $\nu_{\rm nc}/\rho_c$, which controls the strength of momentum exchange between ions and neutrals.
    }
    \label{tab:coupling_regimes}
\end{table*}

\section{Verification of the linear theory}\label{sec:linear_verif}

The aim of this section is to verify that our numerical setup correctly reproduces the predictions of the linear Rayleigh–Taylor theory in all configurations relevant to this work (hydrodynamic, single-fluid MHD, and bi-fluid). We proceed gradually, beginning with the definition and validation of the diagnostic used to measure the instability growth, and then increasing the physical complexity step by step. The linear theory developed in \citetalias{Callies2025} relied on the simplifying assumption that the ion--neutral and neutral--ion collision frequencies were identical on both sides of the density interface. While this choice allowed for a compact analytical formulation, it is not strictly consistent with the numerical implementation used in the present work, where the conserved quantity is the drag coefficient rather than the collision frequency itself. As a result, the collision frequencies naturally vary across the interface in proportion to the local density contrast. In order to ensure a physically consistent comparison between the analytical predictions and the numerical simulations, we relax this assumption and allow the collision frequencies to differ on either side of the interface. We show in Appendix~\ref{app:paper1_corrigendum} that this modification does not introduce any new qualitative regime of instability. Instead, it primarily results in a renormalisation of the effective coupling strength at intermediate wavelengths, while leaving the uncoupled and fully coupled asymptotic limits unchanged. The analytical framework of \citetalias{Callies2025} therefore remains fully applicable, provided that the coupling strength is interpreted in terms of an effective collision frequency.
\subsection{Hydrodynamic reference: compressibility check}
\label{subsec:hydro_compressible}
The theoretical predictions used throughout this work are based on the incompressible linear analysis of \citetalias{Callies2025}, while the simulations solve the fully compressible equations. It is therefore necessary to assess whether compressibility may significantly alter the linear growth rates in the parameter regime explored here. Following \citet{Diaz2012}, a convenient estimate is provided by the dimensionless ratio $gL/c_s^2$, where $L$ is the characteristic perturbation wavelength and $c_s$ the sound speed. When $gL/c_s^2\ll1$, pressure adjusts rapidly compared to buoyancy--driven motions, and the instability behaves effectively incompressibly. Using the largest wavelength present in our simulations, we obtain the conservative estimate
\begin{equation}
    \frac{g_cL}{c_s^2}\simeq 0.1,
    \label{eq:gL_cs2_value}
\end{equation}
indicating that compressibility effects should remain weak during the linear stage. This expectation is confirmed quantitatively by comparing incompressible and compressible hydrodynamic growth rates for a representative wavelength $L_x/4$,
\begin{equation}
    \omega_{\rm th,comp}=6.94\times10^{-4}\ {\rm s^{-1}},\qquad
    \omega_{\rm th,incomp}=7.02\times10^{-4}\ {\rm s^{-1}},
\end{equation}
which differ by less than $2\%$. In the remainder of this work, the incompressible dispersion relation is therefore used as a reference framework to interpret the simulations. We emphasize that it is not expected to predict the detailed temporal evolution of the compressible system at all times, but rather to capture the correct ordering of growth rates and trends across wavelengths and coupling regimes. Small temporal offsets between theoretical predictions and numerical measurements are thus expected and will be accounted for in the analysis.

\subsection{Diagnostic tool: finger--bubble height in the linear and weakly nonlinear regimes}
\label{subsec:diagnostic}

In order to quantify the evolution of the Rayleigh--Taylor instability in a way that is robust and comparable across all physical configurations explored in this work (HD, MHD, and bi--fluid), we follow the vertical penetration of both the rising light--fluid bubbles and the descending heavy--fluid spikes (fingers). Our primary observable is the horizontally averaged heavy--fluid fraction as a function of height,
\begin{equation}
    \langle f_h\rangle(z,t)=\frac{1}{L_x}\int_{0}^{L_x} f_h(x,z,t)\,dx,
    \label{eq:fh_def}
\end{equation}
where $f_h$ is the local mass fraction of heavy fluid. From $\langle f_h\rangle(z,t)$ we define two characteristic interface positions: the finger position $z_f(t)$ as the lowest height in the upper half for which the heavy fluid is still present, and the bubble position $z_b(t)$ as the highest height in the lower half for which light fluid is still present, using fixed thresholds $f_{\rm low}$ and $f_{\rm high}$ (here $f_{\rm low}=0.01$ and $f_{\rm high}=0.99$). In practice, $z_f(t)$ is obtained from the crossings of $\langle f_h\rangle(z,t)=f_{\rm low}$ on the $z>0$ side (spike penetration), while $z_b(t)$ is obtained from the crossings of $\langle f_h\rangle(z,t)=f_{\rm high}$ on the $z<0$ side (bubble rise). We then define the finger--bubble height
\begin{equation}
    h(t)\equiv z_f(t)-z_b(t),
    \label{eq:h_def}
\end{equation}
which directly measures the thickness of the developing mixing region in a way that is insensitive to a drift of the mean interface position. This diagnostic is widely used in RT studies (e.g. \citealt{Stone_2007}) .

\subsection{Linear validation with multiple injected wavelengths}
\label{subsec:linear_validation_multik}

We now validate the linear regime of the instability in a controlled multi--wavelength setting, both in the single--fluid MHD limit and in the bi--fluid framework, before turning to nonlinear and multi--scale effects. We independently inject several wavelengths and analyse their early--time evolution with the diagnostic of Sect.~\ref{subsec:diagnostic}. In all simulations, the initial interface displacement is prescribed with a controlled amplitude proportional to the injected wavelength, $h_0=\varepsilon\lambda$ with  (here $\epsilon\simeq 1/3$), so that all runs start from the same dimensionless configuration. We analyse the normalised height $h(t)/h_0$ and normalise time by the theoretical linear growth rate $\omega_{\rm th}$ associated with each wavelength, defining $\bar{t}=\omega_{\rm th}t$. In the linear regime, correct theory/diagnostic behaviour therefore implies $h/h_0\propto \exp(\bar{t})$ and, crucially, collapse of curves across different wavelengths when plotted as a function of $\bar{t}$.

\begin{figure}
\centering

\begin{subfigure}{\linewidth}
\centering
\includegraphics[width=\linewidth]{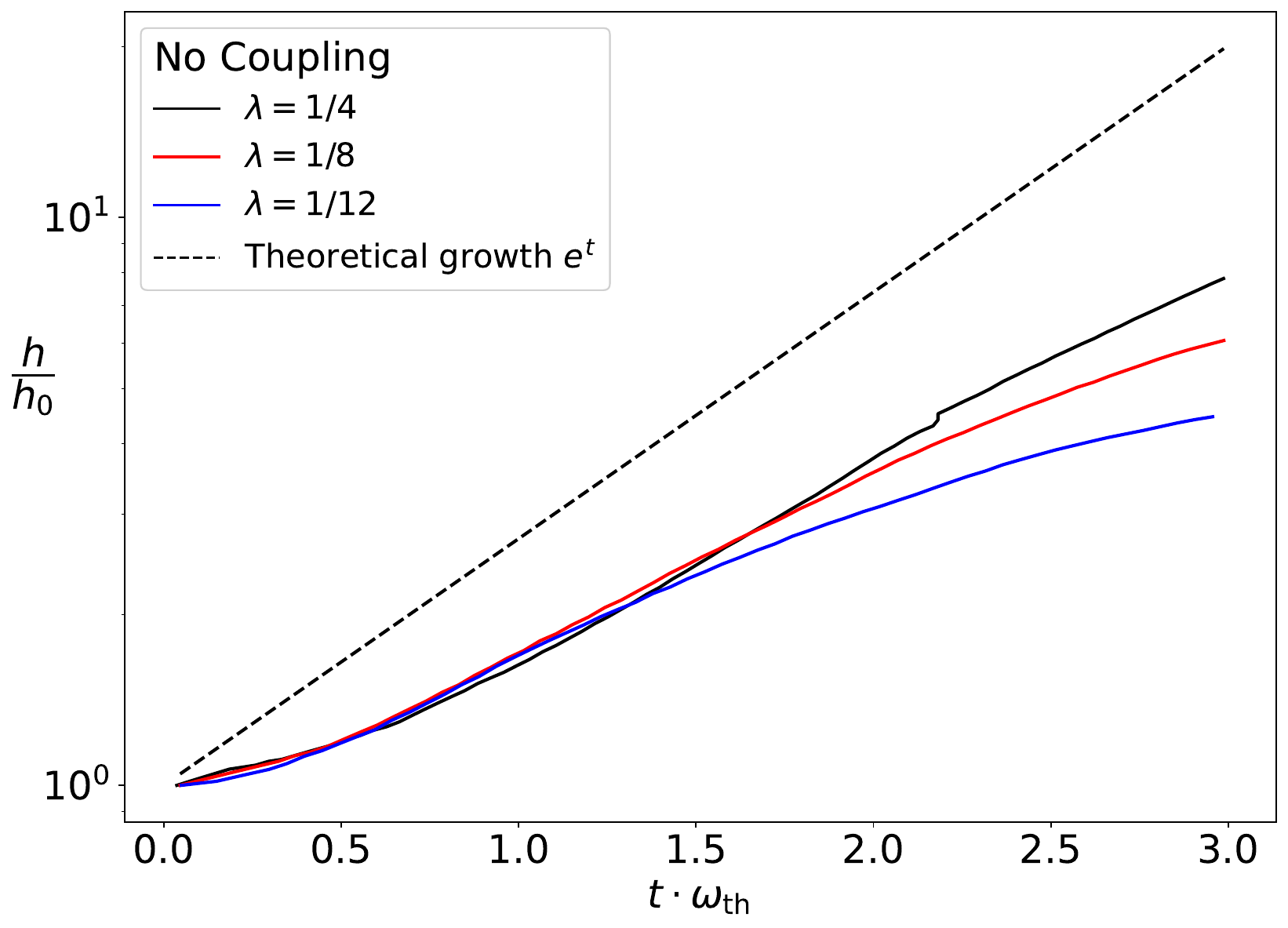}
\caption{NC}
\label{fig:MHD_a0_height_norm}
\end{subfigure}

\medskip

\begin{subfigure}{\linewidth}
\centering
\includegraphics[width=\linewidth]{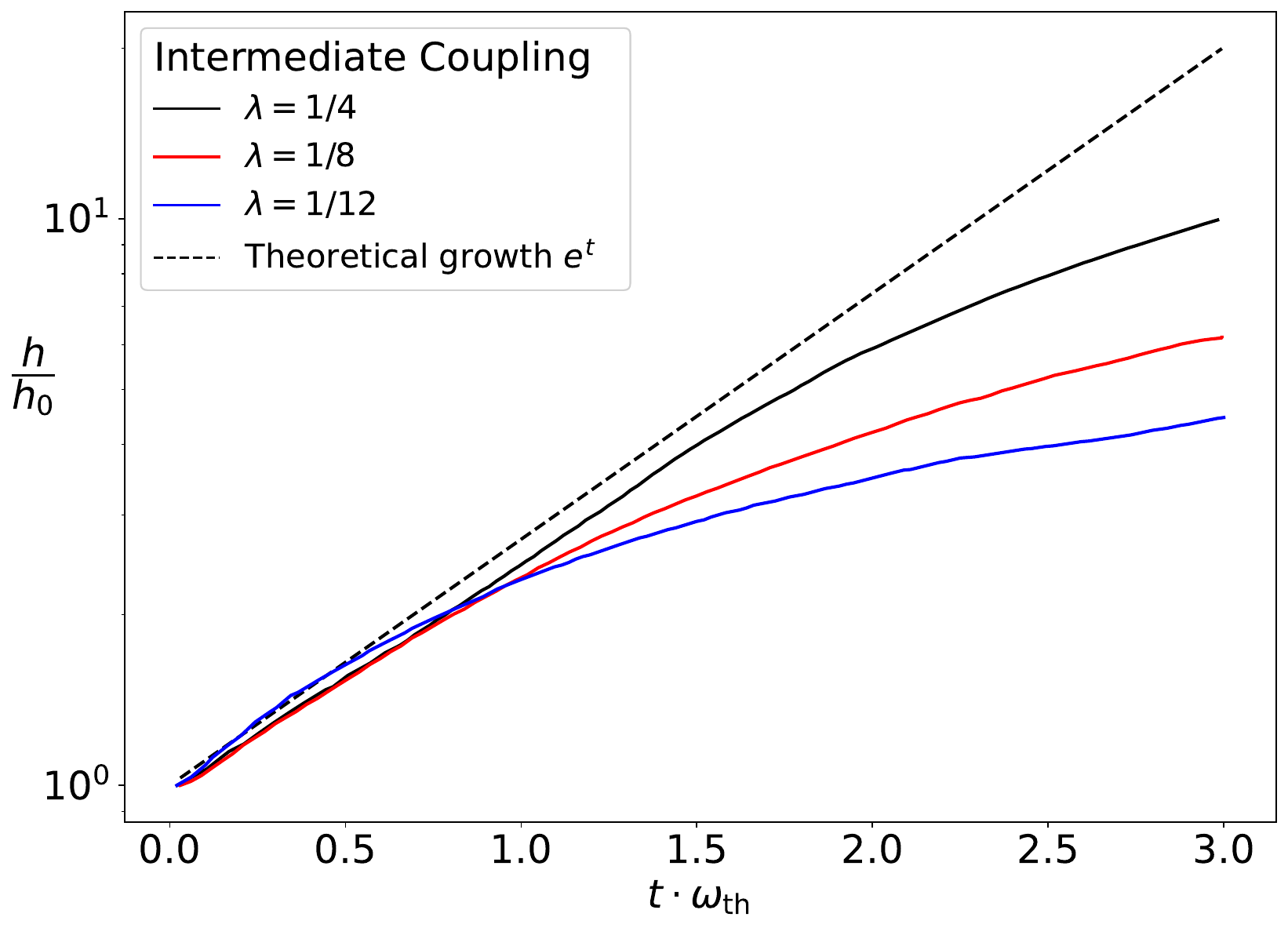}
\caption{IC}
\label{fig:bifluid_growth_comparison}
\end{subfigure}

\caption{
Normalized finger--bubble height $h/h_0$ plotted as a function of the reduced time $\bar{t}=\omega_{\rm th} t$ for three injected wavelengths.  curves correspond to the numerical results, while the solid green line show the corresponding theoretical linear growth rates.
Top panel: no--coupling (NC) case.
Bottom panel: bi--fluid simulation in the intermediate--coupling (IC) regime.}

\end{figure}

We first consider the single--fluid MHD configuration obtained by setting the neutral--ion collision frequency to zero ($\nu_{\rm nc}=0$), with a uniform magnetic field and plasma beta $\beta=3\times10^{2}$. Three wavelengths are selected so as to sample distinct regions of the theoretical growth--rate curve, and each wavelength is injected individually in otherwise identical runs. The resulting evolution of $h/h_0$ is shown in Fig.~\ref{fig:MHD_a0_height_norm}. In the linear regime ($0.5\lesssim \bar{t}\lesssim2$), the three curves collapse and appear as straight lines in semi--logarithmic representation, demonstrating that the predicted exponential growth is accurately recovered once each wavelength is normalised by its corresponding $\omega_{\rm th}$. This validates both the single--fluid dispersion relation and the consistency of the height diagnostic across spatial scales. At later times, the curves naturally deviate from a common exponential as nonlinear effects develop; this separation is expected and simply reflects the fact that shorter wavelengths dominate early growth while longer wavelengths become increasingly influential as the interface displacement becomes comparable to the mode scale.

We then perform the same validation in the bi--fluid model in a representative intermediate--coupling regime (IC), where ion--neutral interactions are significant but do not trivially enforce a single effective fluid. The same three wavelengths are injected with the same dimensionless initial amplitude $h_0/\lambda$, and time is normalised by the theoretical bi--fluid growth rate $\omega_{\rm th}$ for each wavelength and coupling strength. Figure~\ref{fig:bifluid_growth_comparison} shows that the curves again collapse tightly in the linear regime, indicating that the bi--fluid dispersion relation correctly predicts the scale dependence of the linear growth and that the numerical coupling implementation reproduces the expected exponential stage. For completeness, we note that in strictly uncoupled bi--fluid runs ($\nu_{\rm nc}=0$) one may observe a small systematic early--time offset even after normalisation by $\omega_{\rm th}$; since the exponential slopes remain correct, this effect is best interpreted as a transient adjustment of the initial conditions when the two fluids evolve independently rather than as an error in the growth rate itself. Importantly, this delay is absent in the IC configuration shown here, suggesting that even moderate coupling suppresses such early relaxation effects. Overall, the multi--wavelength collapse in both figures provides a stringent validation of the linear theory and of the diagnostic pipeline used throughout the remainder of the paper. 

The results presented in this section demonstrate that the linear growth of the instability in our simulations is accurately captured by the incompressible dispersion relation derived in \citetalias{Callies2025}, within the expected limitations associated with compressibility and finite-amplitude initialization. Growth rates, mode ordering, and magnetic stabilization effects are consistently recovered across wavelengths and coupling regimes. The small discrepancies observed at early times are attributable to compressibility-induced delays and diagnostic uncertainties, rather than to a breakdown of the theoretical framework. This validates both the numerical setup and the analysis tools employed in the remainder of this work, and ensures that the deviations discussed in the following sections genuinely arise from nonlinear and ambipolar diffusion effects.

\section{Non Linear Analysis}\label{sec:NonLin}
We now examine the nonlinear evolution of the Rayleigh--Taylor instability and the role of ambipolar diffusion beyond the linear regime. The analysis starts from single--mode perturbations to isolate nonlinear effects, and is then extended to multi--wavelength configurations in both hydrodynamic and magnetized regimes. 

\subsection{One wavelength}

In the nonlinear stage of the Rayleigh--Taylor instability, the growth of the mixing layer is often described by the quadratic law \citep{Sharp1984,D04_article}
\begin{equation}
h(t) \;\approx\; \alpha\,\mathcal{A}\,g\,t^{2},
\label{eq:t2_law}
\end{equation}
where $\mathcal{A}$ is the Atwood number, $g$ the gravitational acceleration, and $\alpha$ a dimensionless coefficient characterizing the effective nonlinear growth. This scaling is commonly associated with a self--similar regime in which the dynamics is governed by a single macroscopic length scale $h(t)$ and a characteristic velocity $U\sim\dot h$, while the detailed flow physics is absorbed into the value of $\alpha$. In the following, this quadratic scaling is therefore used as a reference framework to assess how ambipolar diffusion modifies the nonlinear evolution, rather than as an a priori universal growth law.

\begin{figure}
    \centering
    \includegraphics[width=0.98\linewidth]{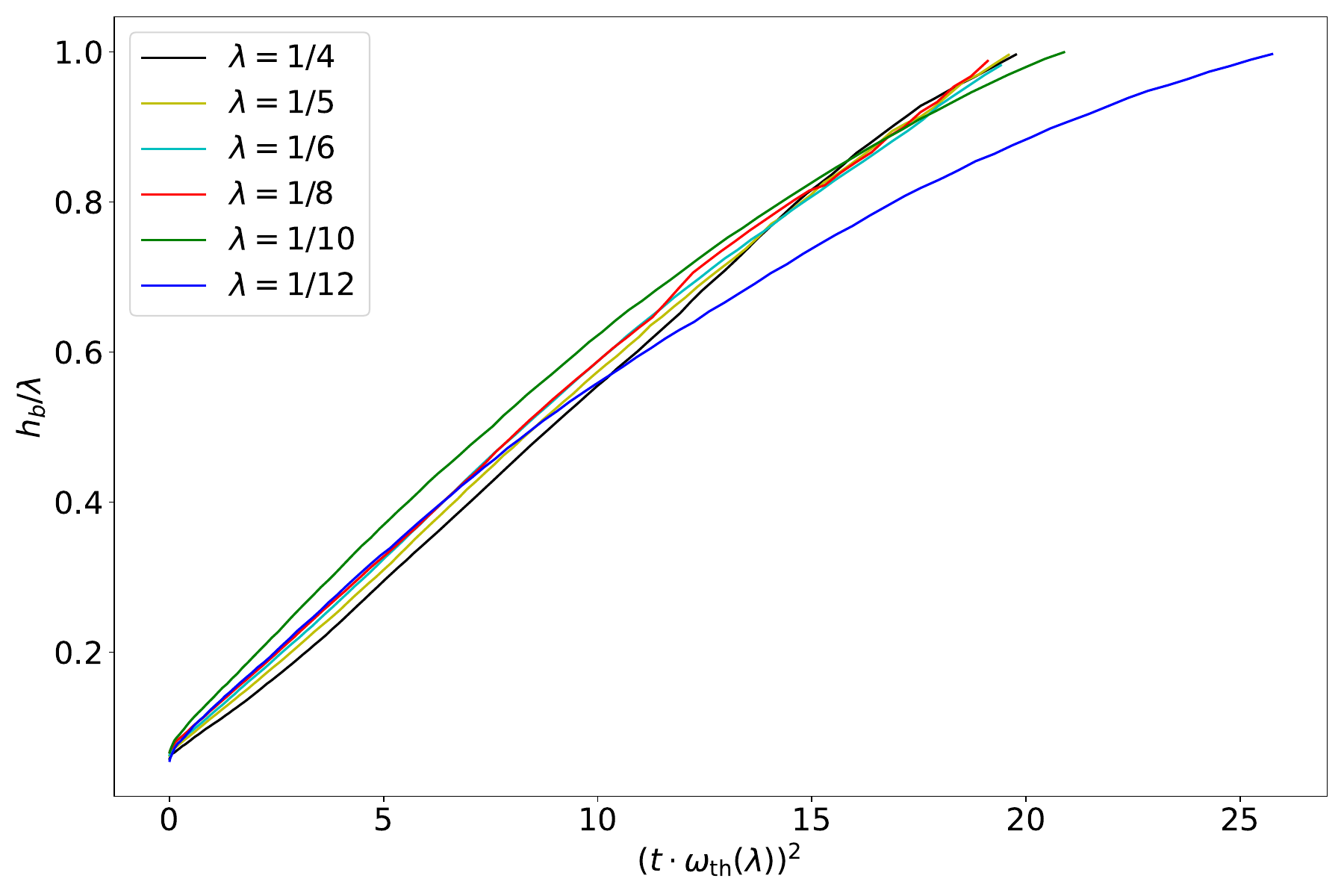}
    \caption{Quadratic scaling of the mixing height in the single–fluid case ($\nu_{\rm nc}=0$). Each color represents a different value of the perturbation wavelength.}
    \label{fig:h_t2_nu0}
\end{figure}

We first consider the uncoupled case ($\nu_{\rm nc}=0$), which provides a reference single--fluid behavior. Figure~\ref{fig:h_t2_nu0} shows the dimensionless mixing height $h/\lambda$ as a function of the rescaled time $t\,\omega_{\rm th}$, where $\omega_{\rm th}$ is the linear growth rate predicted by the two--fluid linear analysis. For all considered wavelengths, the curves are approximately linear, confirming the quadratic growth of the mixing layer, and their slopes remain comparable over a broad range of wavelengths, in agreement with previous numerical studies. At the shortest wavelengths, a tendency toward saturation is observed at small scales, which can be attributed to enhanced dissipation and finite resolution effects. Overall, in the absence of coupling, the quadratic model provides an accurate and robust description of the nonlinear growth in our 2D simulations. Before turning to the coupled regimes, it is important to stress that the quadratic model in Eq~(\ref{eq:t2_law}) should be regarded as an effective description rather than a strict law, especially in two--dimensional configurations. As emphasized by \citet{Kalluri2025}, self--similarity in Rayleigh--Taylor mixing does not necessarily imply an exactly quadratic growth at all times, but rather the emergence of a regime controlled by a single macroscopic length scale, in which suitably defined dimensionless coefficients remain approximately constant. In 2D, the coexistence of direct and inverse cascades and the persistence of large--scale coherent structures make this regime intrinsically less robust than in three dimensions, so that temporal modulations of the effective growth rate are expected even in the absence of additional physics. In this context, the parameter $\alpha$ should be interpreted as a time--averaged measure of the nonlinear growth efficiency, whose applicability may be limited when additional mechanisms interfere with the inertial--buoyancy balance.

\begin{figure}
\begin{subfigure}{\linewidth}
\centering
\includegraphics[width=0.98\linewidth]{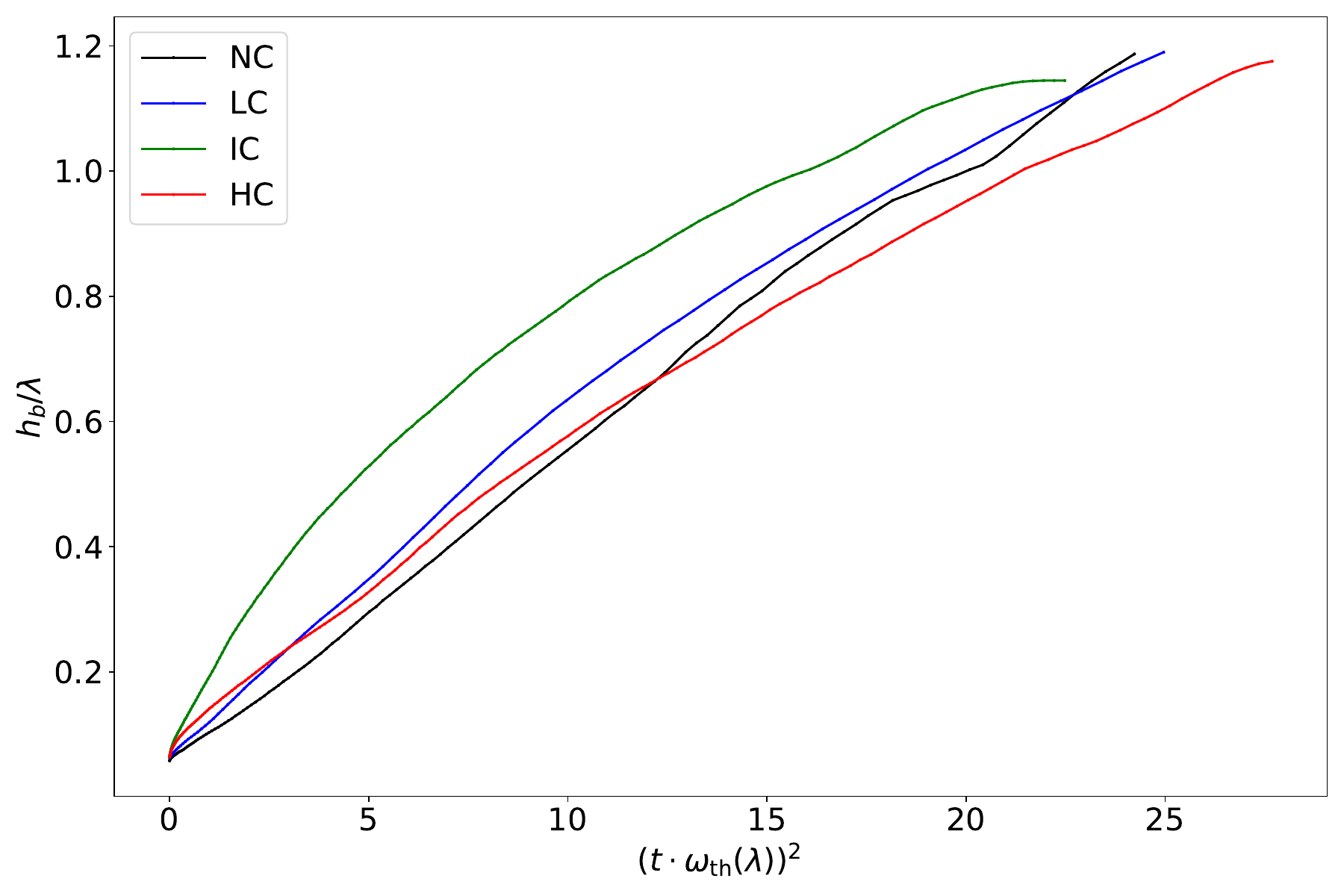}
\caption{Simulation}
\label{fig:h_t2_var_nu}
\end{subfigure}

\medskip

\begin{subfigure}{\linewidth}
\centering
    \includegraphics[width=\linewidth]{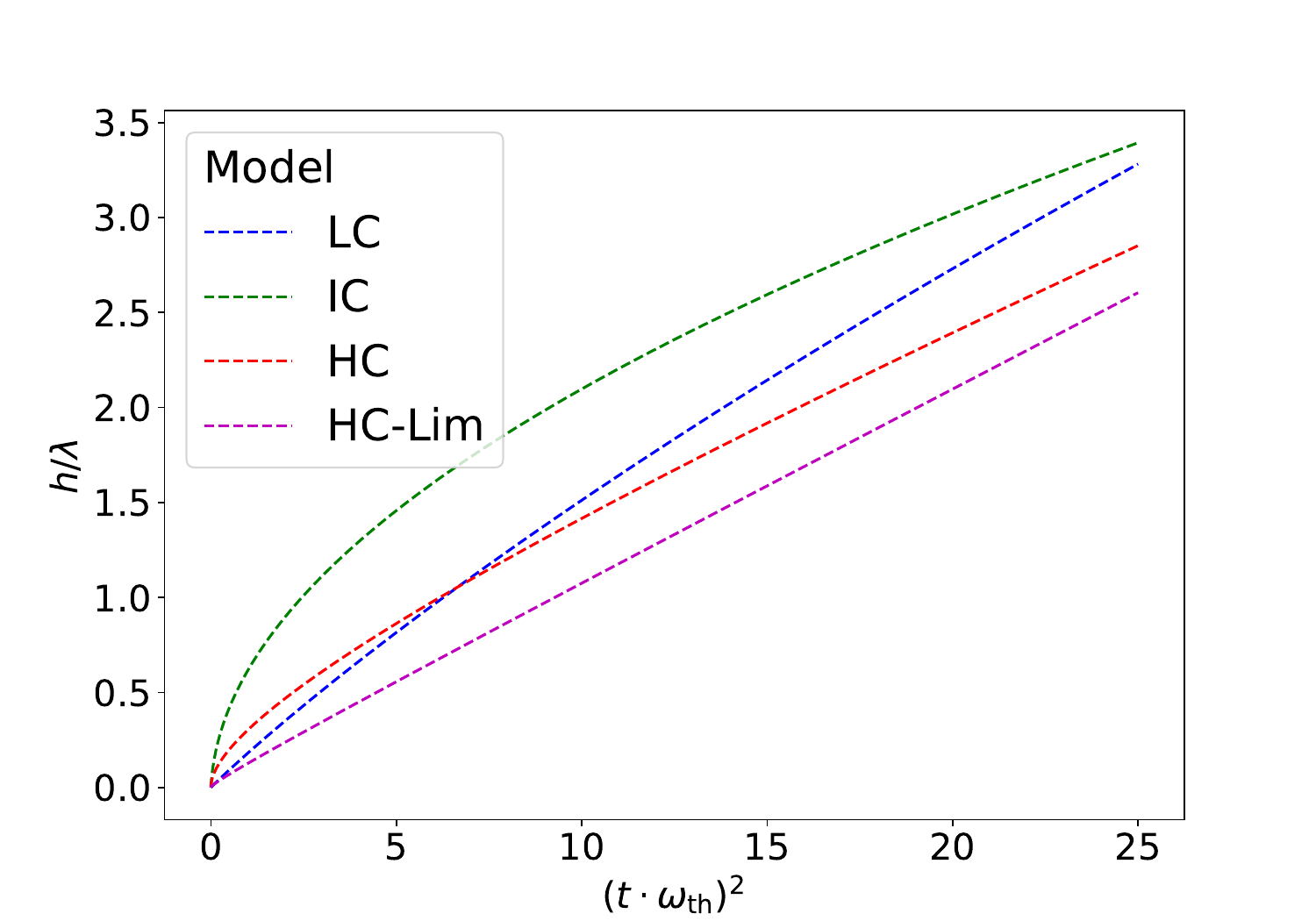}
    \caption{Model}

    \label{fig:toy_model}
\end{subfigure}
\caption{Quadratic scaling of the mixing height for No , Low, Intermediate , High Coupling cases (NC, LC, IC, HC) for $\lambda=1/4 L_x$. Top Panel: simulation result, bottom panel : Theoretical evolution, predicted by the local two--fluid simplified model}
\end{figure}

When ambipolar coupling is introduced, the nonlinear evolution departs more profoundly from the quadratic reference. As shown in Fig.~\ref{fig:h_t2_var_nu}, the ambipolar regime is characterized by a pronounced curvature of the growth curves, indicating a sub--quadratic evolution over an extended time interval. The mixing height therefore grows more slowly than predicted by Eq.~(\ref{eq:t2_law}), reflecting a strong reduction of the instantaneous growth rate compared to the uncoupled case. At later times, the curves tend to straighten again, suggesting a gradual recovery of a quasi--quadratic behavior, but with a significantly reduced effective coefficient. This behaviour shows that ambipolar diffusion does not simply renormalize the quadratic law through a smaller constant, but instead induces a genuinely time--dependent growth process. The failure of the quadratic description in the ambipolar regime can be understood from a simple physical argument. In the classical single--fluid picture, the buoyancy power injected into the flow is assumed to be directly converted into bulk kinetic energy at the scale of the mixing layer, so that the nonlinear dynamics can be closed in terms of the single macroscopic length scale $h(t)$. In a partially ionized medium, however, buoyancy initially accelerates only the charged component, while neutrals are entrained progressively through ion--neutral collisions. During this phase, a finite slip velocity develops and a non--negligible fraction of the injected buoyancy power is transferred into relative ion--neutral motion and dissipated by drag on a characteristic coupling timescale. As a result, the instantaneous growth rate of the mixing layer depends not only on $h(t)$, but also on the time--dependent state of coupling between the two fluids, preventing the closure of the growth law in terms of $h$ alone.

To interpret the departure from a purely quadratic growth observed in Fig.~\ref{fig:h_t2_var_nu}, we consider a minimal local two--fluid simplified model in which buoyancy acts only on the charged component, while ions and neutrals exchange momentum through linear ion--neutral drag. The model assumes spatially averaged velocities, a constant density ratio $\chi\equiv\rho_c/\rho_n$, and neglects pressure gradients, turbulent transport, and spatial intermittency. Its purpose is not to provide a quantitative description of the nonlinear Rayleigh--Taylor instability, but to isolate the dynamical consequences of time--dependent ion--neutral coupling.

The evolution of the charged and neutral velocities is governed by
\begin{align}
\rho_c \frac{dU_c}{dt} &= \mathcal{A}g\,\rho_c - \rho_c \nu_{\rm nc}(U_c-U_n),\\
\rho_n \frac{dU_n}{dt} &=  \rho_c \nu_{\rm nc}(U_c-U_n),
\end{align}
where $\nu_{\rm nc}$ is the neutral--on--charge collision frequency and $\Delta u\equiv U_c-U_n$ is the slip velocity. Subtracting the two equations yields a closed evolution equation for the drift:
\begin{align}
    \frac{d\Delta u}{dt}&=\mathcal{A}g-\nu_{\rm nc}(1+\chi)\Delta u ,
\end{align}
which relaxes exponentially on the coupling timescale $\tau_c=[\nu_{\rm nc}(1+\chi)]^{-1}$. Summing the momentum equations gives the total momentum balance,
\begin{align}
    \frac{d}{dt}(\rho_c U_c+\rho_n U_n)&=\mathcal{A}g\,\rho_c\,
\end{align}
 from which the center--of--mass velocity grows linearly as $U=\mathcal{A}g(\rho_c/\rho_{\rm tot})\,t$.

Combining these two results, the charged velocity can be written explicitly as the sum of a center--of--mass contribution and a transient drift term. Identifying the charged bulk velocity with the growth rate of the mixing layer, $U_c\simeq\dot h$, one obtains an analytical expression for the mixing height,
\begin{equation}
h(t)-h_0=\frac{\mathcal{A}g\,\chi}{2(1+\chi)}\,t^2+\frac{\mathcal{A}g}{\nu_{\rm nc}^2(1+\chi)^3}\left[(1+\chi)\nu_{\rm nc}t-\bigl(1-e^{-(1+\chi)\nu_{\rm nc}t}\bigr)\right].
\end{equation}
Introducing the reduced time $\tilde t\equiv\nu_{\rm nc}t$, the mixing height can be rewritten in dimensionless form as
\begin{equation}
h(\tilde t/\nu_{\rm nc})-h_0
=
\frac{\mathcal{A}g}{\nu_{\rm nc}^2}
\left[
\frac{\chi}{2(1+\chi)}\,\tilde t^{\,2}
+
\frac{1}{(1+\chi)^3}
\left(
(1+\chi)\tilde t-\bigl(1-e^{-(1+\chi)\tilde t}\bigr)
\right)
\right].
\end{equation}
In this form, the growth law depends explicitly only on the reduced elapsed time $\nu_{\rm nc}t$, making clear that the nonlinear evolution is governed by the relative ordering of the buoyancy timescale and the ion--neutral coupling time.

At early times, $\nu_{\rm nc}t\ll1$, ions and neutrals are effectively decoupled and the charged component accelerates freely, recovering the classical quadratic growth $h\propto\mathcal{A}g\,t^2$. In the opposite limit, $\nu_{\rm nc}t\gg1$, ions and neutrals are locked together and the system behaves as a single fluid of total inertia, yielding $h\propto\mathcal{A}g(\rho_c/\rho_{\rm tot})\,t^2$. The intermediate regime $\nu_{\rm nc}t\sim1$ corresponds to a transient phase during which neutrals are progressively entrained by drag, leading to a time--dependent increase of the effective inertia and a reduced instantaneous growth rate.

This behaviour is illustrated in Fig.~\ref{fig:toy_model}, where the theoretical prediction for $h/\lambda$ is plotted as a function of $(t\cdot\omega_{\rm th})^2$ for weakly (LC), intermediately (IC), and strongly (HC) coupled cases. The model reproduces the pronounced curvature of the IC curve, as well as the reduced asymptotic prefactor in the HC limit, in close qualitative agreement with the simulations. The remaining offset in the normalization of $h/\lambda$ reflects the absence of a phenomenological efficiency factor (e.g. $\alpha\,\mathcal{A}g$ as introduced in classical self--similar RT models), the present approach being intended solely as a minimal dynamical interpretation.

Importantly, the intermediate--coupling regime does not represent a breakdown of self--similar behaviour, but rather a modified form of self--similarity in which the growth law contains both quadratic and linear contributions. The linear term reflects the finite time required for momentum transfer between the two fluids and the associated evolution of the effective inertia of the mixing layer. The observed sub--quadratic growth therefore arises naturally from the intrinsic two--fluid dynamics, without invoking additional nonlinear saturation mechanisms.

\subsection{Multi-wavelength regime: hydrodynamic reference}
\label{sec:nonlinear_multi}

We now extend the nonlinear analysis to multi--wavelength initial conditions, in which the interface is perturbed over a broad range of spatial scales. We first consider the hydrodynamic configurations as a reference for the nonlinear evolution under multi--wavelength perturbations. All comparisons are performed at an equivalent nonlinear stage defined by a fixed normalized mixing layer thickness, $\Delta h/L_x \simeq 0.35$.

\begin{figure*}
\centering
\includegraphics[width=0.98\linewidth]{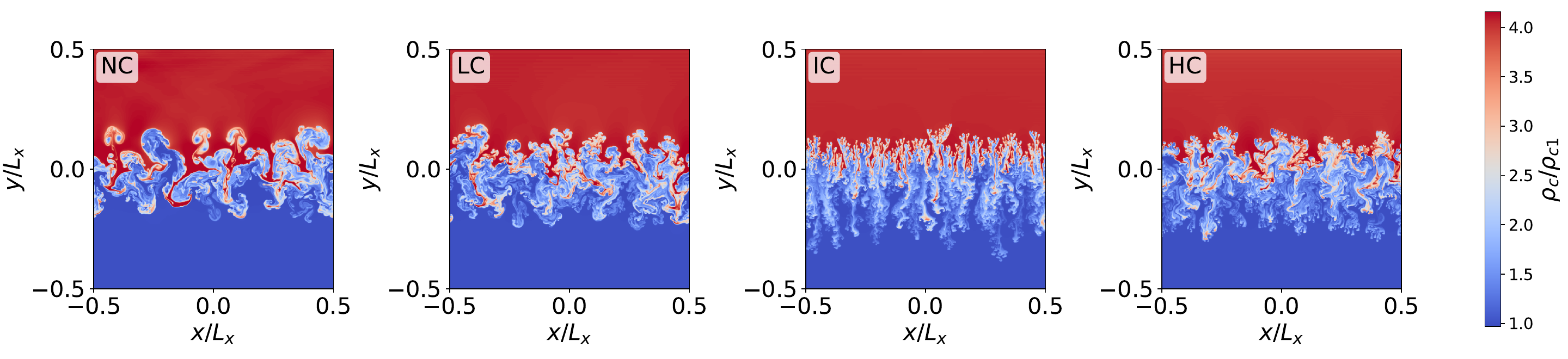}
\caption{
Snapshots of the charge density $\rho_c$ in hydrodynamic simulations with multi--wavelength initial perturbations, extracted at an equivalent nonlinear stage defined by a fixed normalized mixing layer thickness, $\Delta h/L_x \simeq 0.35$.
From left to right: increasing coupling strength ($\alpha=0$, weak, intermediate, and strong coupling).
All snapshots are shown using the same spatial window and color scale. (the density of the charge fluid $\rho_c$ is given in unit of the $\rho_{c1}$, the lighter fluid)}
\label{fig:hd_snapshots_dh}
\end{figure*}

Figure~\ref{fig:hd_snapshots_dh} illustrates the nonlinear morphology of the mixing layer in the hydrodynamic multi--wavelength simulations at a fixed normalized thickness, $\Delta h/L_x \simeq 0.35$ , it shows snapshots of the charge density $\rho_c$ for increasing coupling strength.  
In the uncoupled case ($\alpha=0$), the interface is dominated by a small number of large--scale structures. The initially broadband perturbation rapidly evolves toward coherent bubbles and fingers that undergo efficient lateral merging, leading to extended plumes and a relatively smooth mixing layer. As the coupling strength increases, the morphology is progressively modified. For weak coupling, the global organization remains close to the uncoupled reference, although small--scale corrugations and secondary structures become more visible along the sides of the plumes. In contrast, the intermediate--coupling regime exhibits a markedly different behavior: at the same mixing height, the interface is significantly more fragmented, with numerous thin fingers and strongly reduced lateral coalescence. Large--scale plumes fail to emerge, and the mixing region retains a highly structured appearance down to smaller spatial scales. In the strongly coupled regime, the interface recovers a smoother and more coherent morphology, with reduced small--scale fragmentation and a renewed tendency toward large--scale organization. To quantify these morphological differences, we compute one--dimensional power spectra of the charge density $\rho_c$ along the horizontal direction $x$, evaluated at the same normalized mixing layer thickness. The spectra are obtained using a Welch--averaged estimator \citep{Welch1967} and normalized by their maximum value to highlight the relative distribution of power across scales.

\begin{figure}
    \centering
    \includegraphics[width=\linewidth]{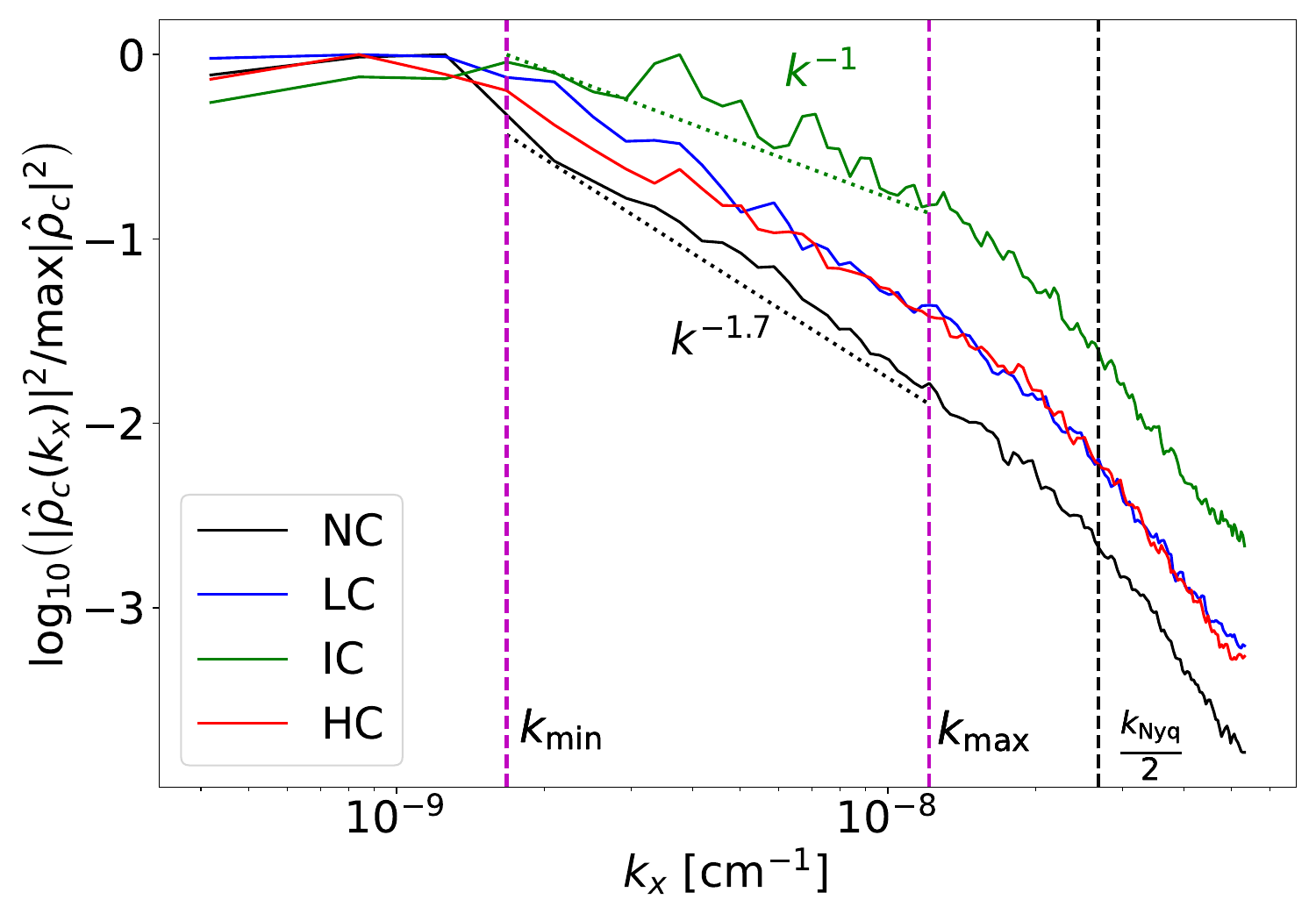}
\caption{ One--dimensional power spectrum of the charge density $\rho_c$ along the horizontal direction $x$, computed for the hydrodynamic multi--wavelength simulations at a fixed nonlinear stage defined by $\Delta h/L_x \simeq 0.35$. The spectra correspond to $P_{\rho_c}(k_x)=\langle|\hat{\rho}_c(k_x)|^2\rangle$ and are normalized by their maximum value. Vertical dashed lines indicate the minimum and maximum injected wavenumbers, as well as half the Nyquist limit.}
\label{fig:hd_spectra_rho}
\end{figure}

Figure~\ref{fig:hd_spectra_rho} shows that the redistribution of structures observed in real space is accompanied by clear, systematic changes in spectral content. In the uncoupled case, the power is predominantly concentrated at small wavenumbers, consistent with the dominance of large--scale plumes. As the coupling strength increases, power is progressively shifted toward larger horizontal wavenumbers (i.e $k_x$), indicating an enhanced contribution from smaller spatial scales at the same mixing height. This redistribution is most pronounced in the intermediate--coupling regime, which exhibits the strongest relative enhancement of high--$k_x$ power. In the strongly coupled case, the spectrum departs less markedly from the uncoupled reference, consistent with the partial recovery of large--scale organization seen in the snapshots.

Taken together, the snapshots and density spectra demonstrate that ion--neutral coupling has a non--monotonic impact on the nonlinear, multi--scale organization of the mixing layer. At fixed $\Delta h/L_x$, intermediate coupling is the most effective at inhibiting large--scale coalescence and maintaining power at smaller horizontal scales, while both weak and strong coupling favor the emergence of larger--scale structures. This hydrodynamic reference establishes that ambipolar coupling can fundamentally reorganize the nonlinear cascade of structures even in the absence of magnetic forces, providing a baseline for interpreting the magnetized configurations discussed below.

\subsection{Multi-wavelength MHD regime}
\begin{figure*}
\centering
\includegraphics[width=0.98\linewidth]{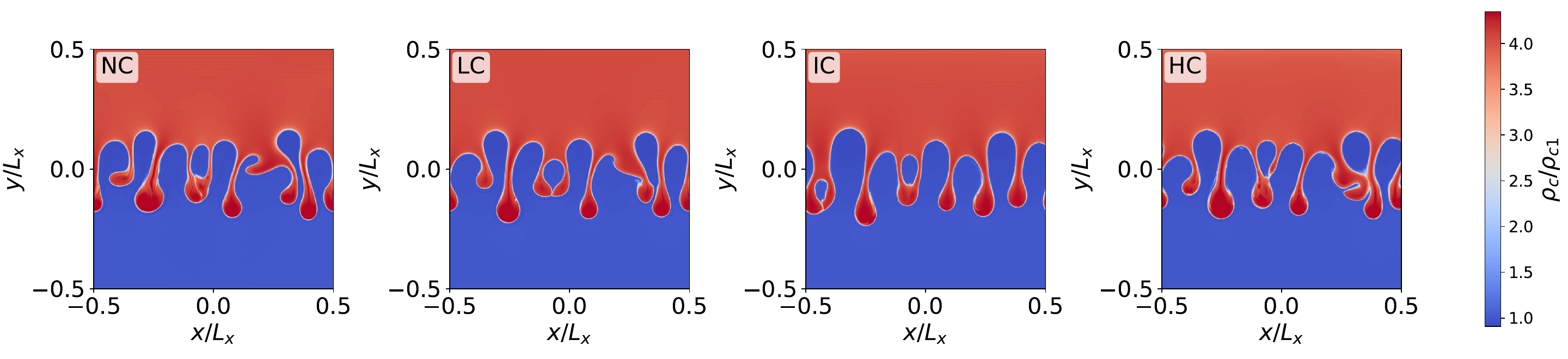}
\caption{
Snapshots of the charge density  $\rho_c$ in MHD simulations with multi--wavelength initial perturbations, extracted at an equivalent nonlinear stage defined by a fixed normalized mixing layer thickness, $\Delta h/L_x \simeq 0.35$.
From left to right, the panels correspond to increasing coupling strength: $\alpha=0$, weak coupling, intermediate coupling, and strong coupling.
All snapshots are shown using the same spatial window and color scale.(the density of the charge fluid $\rho_c$ is given in unit of the $\rho_{c1}$, the lighter fluid)
}
\label{fig:mhd_snapshots_dh}
\end{figure*}

We now shift to 2D MHD simulations with multi wavelength. Figure~\ref{fig:mhd_snapshots_dh} shows snapshots of the charge density $\rho_c$ for the magnetized simulations at an equivalent nonlinear stage, $\Delta h/L_x \simeq 0.35$, for increasing values of the coupling parameter
$\alpha$.

At $\Delta h/L_x \simeq 0.5$, all magnetized simulations exhibit a fully developed mixing layer whose overall morphology differs markedly from the hydrodynamic reference at the same nonlinear stage. In all MHD cases, the interface appears globally smoother, with a reduced level of small--scale corrugations and fewer secondary structures, consistent with the suppression of short--wavelength features by magnetic tension. In the uncoupled MHD case ($\alpha=0$), the mixing layer is organized into well--defined bubbles and fingers, but the interface retains a noticeable degree of fine structure. Small protrusions and lateral modulations are present along the sides of the main plumes, indicating that, in the absence of ion--neutral coupling, the charged component can still sustain a significant level of small--scale variability despite the presence of the magnetic field. For weak coupling, the overall appearance of the mixing layer remains similar, but the interface becomes slightly more regular. The fingers are more uniform in shape, and the smallest--scale features observed in the uncoupled case are less frequent, suggesting a partial damping of secondary structures as the coupling increases. In the intermediate--coupling regime ($\alpha=7.5\times10^{20}$), the morphology becomes noticeably more coherent. The dominant bubbles are comparatively straight and elongated, with smooth lateral boundaries and very limited internal substructure. The interface is dominated by a small number of clean, well--defined plumes, and fine--scale corrugations are strongly reduced. In the strongly coupled case, the mixing layer remains globally similar in extent, but the interface recovers a degree of complexity. While large--scale bubbles persist, thin fingers and localized distortions reappear along the interface, leading to a more intricate and less uniform morphology than in the intermediate case. This indicates that the influence of coupling on the nonlinear organization of the flow is not monotonic with $\alpha$.

\begin{figure}
    \centering
    \includegraphics[width=\linewidth]{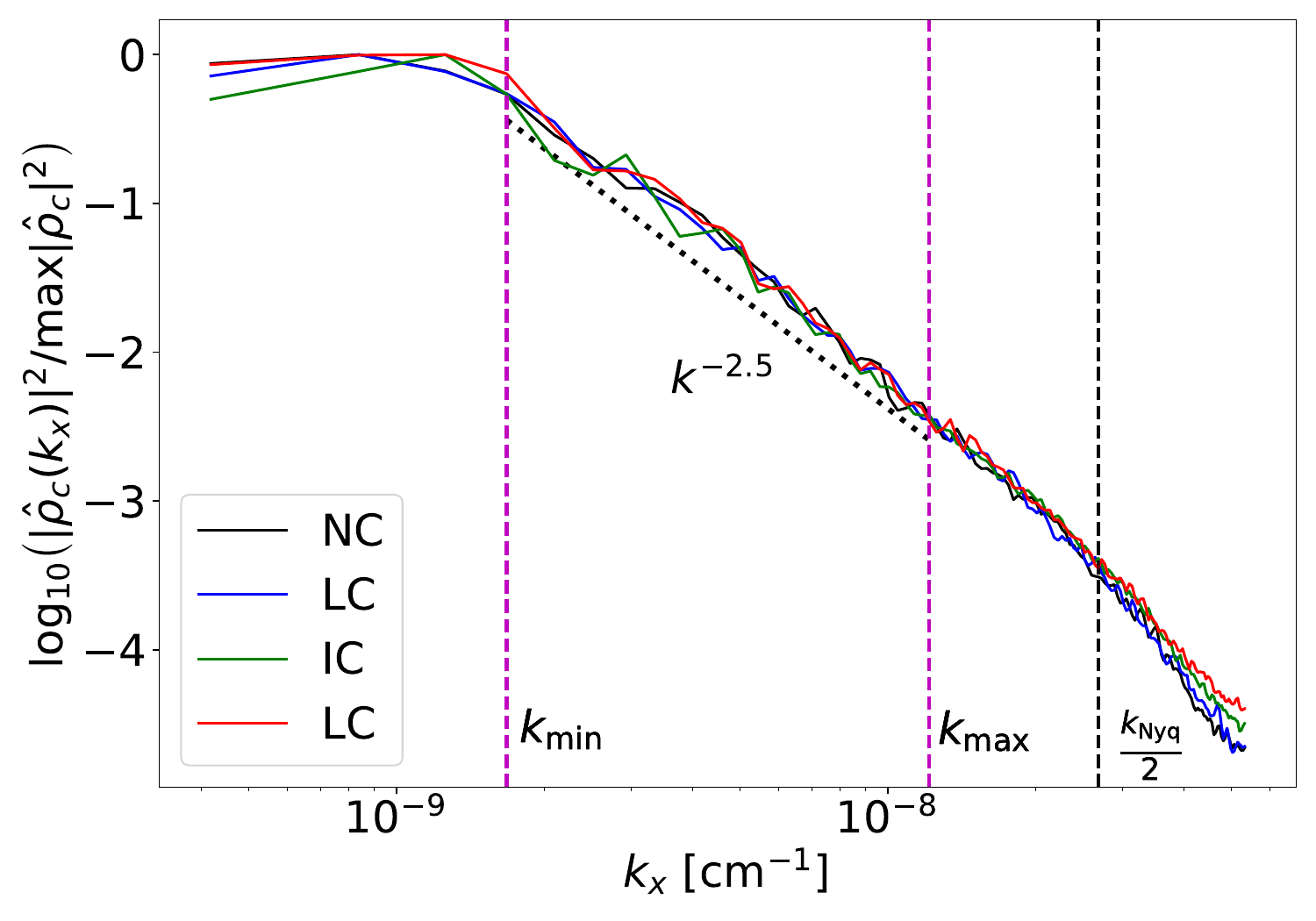}
 \caption{ One--dimensional power spectrum of the charge density $\rho_c$ along the horizontal direction $x$, computed for the magnetized multi--wavelength simulations at an equivalent nonlinear stage defined by $\Delta h/L_x \simeq 0.35$. All spectra are obtained using the same spectral procedure and are normalized by their maximum value. The different curves correspond to increasing ion--neutral coupling strength, labelled as NC, LC, IC, and HC (see Table~3). Vertical dashed lines indicate the minimum and maximum injected wavenumbers, as well as half the Nyquist limit. }
\label{fig:mhd_spectra_rho}

\end{figure}

To further quantify the spatial organization of the flow, we compute the one--dimensional power spectra of the charge density along the horizontal direction at the same nonlinear stage, $\Delta h/L_x \simeq 0.35$. The spectra, shown in Fig.~\ref{fig:mhd_spectra_rho}, are obtained using a Welch--averaged estimator applied within the mixing layer. Despite the morphological differences observed in real space, the spectra exhibit no significant dependence on the coupling strength over the range of resolved scales. In particular, both the spectral slope and the relative distribution of power across horizontal wavenumbers remain remarkably similar between the different MHD runs. This indicates that, at this advanced stage of the nonlinear evolution, the global statistical distribution of power in the charge density is largely insensitive to the value of $\alpha$, even though the geometry of individual bubbles and fingers continues to vary.

\begin{figure*}
\centering
\includegraphics[width=\linewidth]{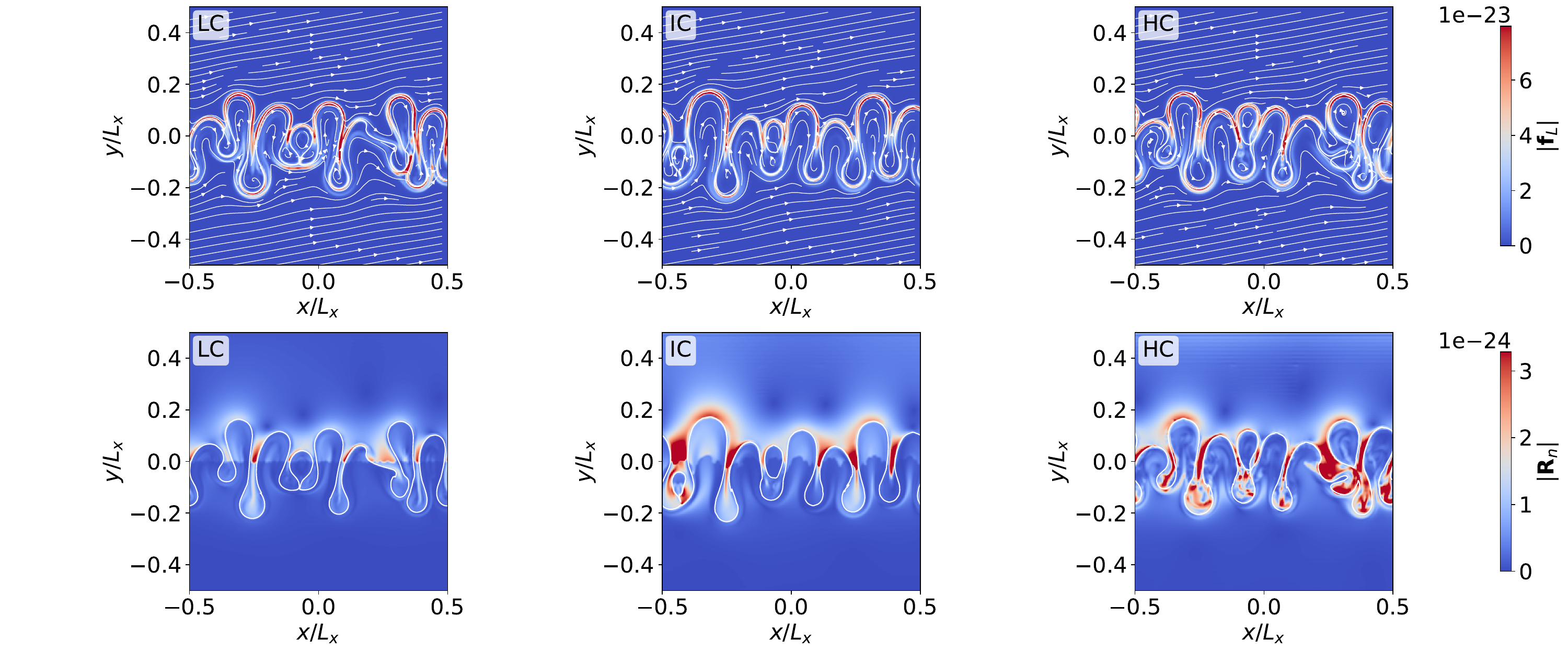}
\caption{Spatial organization of force densities in the magnetized multi--wavelength simulations at a fixed nonlinear stage $\Delta h/L_x \simeq 0.35$. Top row: magnitude of the Lorentz force acting on the charged fluid, $|\mathbf{f}_L|=|\mathbf{J}\times\mathbf{B}|/4\pi$, with magnetic field lines overlaid. Bottom row: magnitude of the ion--neutral drag force, $|\mathbf{R}_n|=\alpha\,\rho_c\rho_n|\mathbf{v}_c-\mathbf{v}_n|$, with streamlines of the relative velocity $\Delta\mathbf{v}=\mathbf{v}_c-\mathbf{v}_n$. White contours delineate the mixing layer. From left to right, the coupling strength increases while the spatial window and color scales are kept identical within each row.}
\label{fig:forces_maps}
\end{figure*}

To identify the physical mechanisms responsible for the nonlinear reorganization of the mixing layer in the magnetized simulations, we first examine spatial maps of the dominant force densities at a fixed nonlinear stage, defined by $\Delta h/L_x \simeq 0.35$. Figure~\ref{fig:forces_maps} displays, for each coupling regime, the Lorentz force acting on the charged component and the ion--neutral drag force acting on the neutrals, computed directly from the instantaneous fields at the selected stage. The top row shows the magnitude of the Lorentz force $\mathbf{f}_L=(\mathbf{J}\times\mathbf{B})/4\pi$, together with magnetic field lines, where $\mathbf{J}=\nabla\times\mathbf{B}$ and all quantities are expressed in cgs units. The Lorentz force is strongly localized along the mixing interface, with enhanced amplitudes around the sides of rising bubbles and sinking fingers, as well as in the ``belly'' regions where magnetic field lines experience the strongest curvature. This spatial organization indicates that magnetic forces primarily act through magnetic tension and pressure associated with field-line bending, rather than as a volume-filling contribution. In particular, the confinement of $\mathbf{f}_L$ to thin layers surrounding the interfaces is consistent with a tension-dominated magnetic response, in which magnetic stresses locally redistribute momentum and energy without producing extended dissipation throughout the bulk of the mixing layer. The bottom row shows the magnitude of the ion--neutral drag force $\mathbf{R}_n=\alpha\,\rho_c\rho_n(\mathbf{v}_c-\mathbf{v}_n)$, together with streamlines of the relative velocity $\Delta\mathbf{v}=\mathbf{v}_c-\mathbf{v}_n$. In contrast with the Lorentz force, the drag contribution occupies a broader fraction of the mixing region and highlights extended zones of strong inter-fluid slip. Enhanced drag is preferentially associated with compressive regions where the density increases, as well as with shear layers around plume edges and, at later stages, near the bases of the fingers. The streamlines of $\Delta\mathbf{v}$ show that ambipolar diffusion acts by driving charges across the magnetic field in these dynamically active regions, producing a spatially heterogeneous pattern of momentum exchange between charges and neutrals. Taken together, these maps reveal a clear separation of roles between the two force channels: magnetic forces primarily act through localized stresses tied to field-line curvature at the interfaces, while ion--neutral drag operates over a wider region of the mixing layer and directly mediates the dissipation of kinetic energy through inter-fluid drift. This spatial complementarity provides a physical basis for understanding how similar global morphologies can arise from different local energy conversion pathways.

\begin{figure}
\centering
\begin{subfigure}{\linewidth}
\centering
\includegraphics[width=\linewidth]{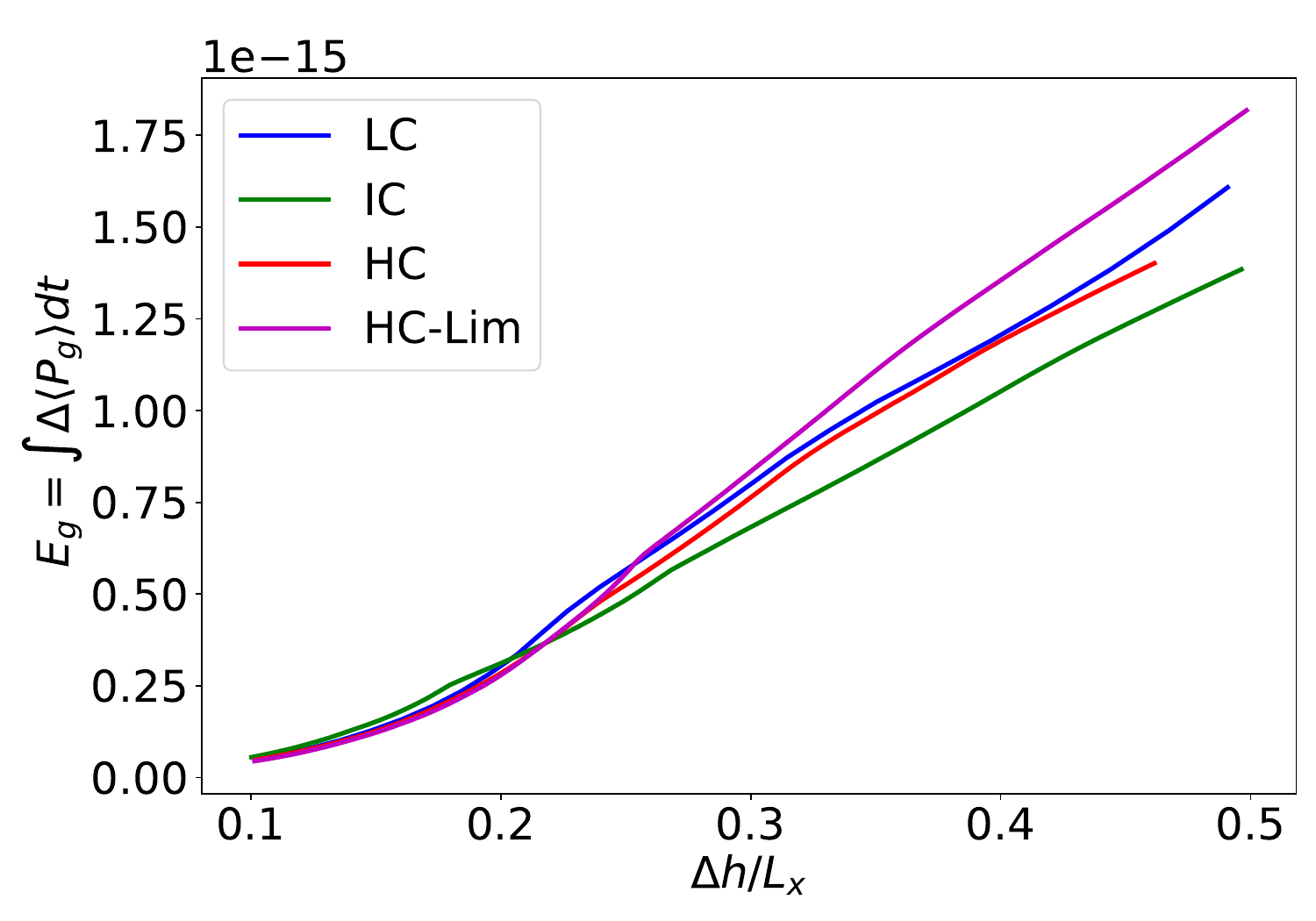}
\caption{Cumulative gravitational energy injection $E_g(t)=\int_0^t\langle P_g\rangle\,dt$.}
\end{subfigure}
\medskip
\begin{subfigure}{\linewidth}
\centering
\includegraphics[width=\linewidth]{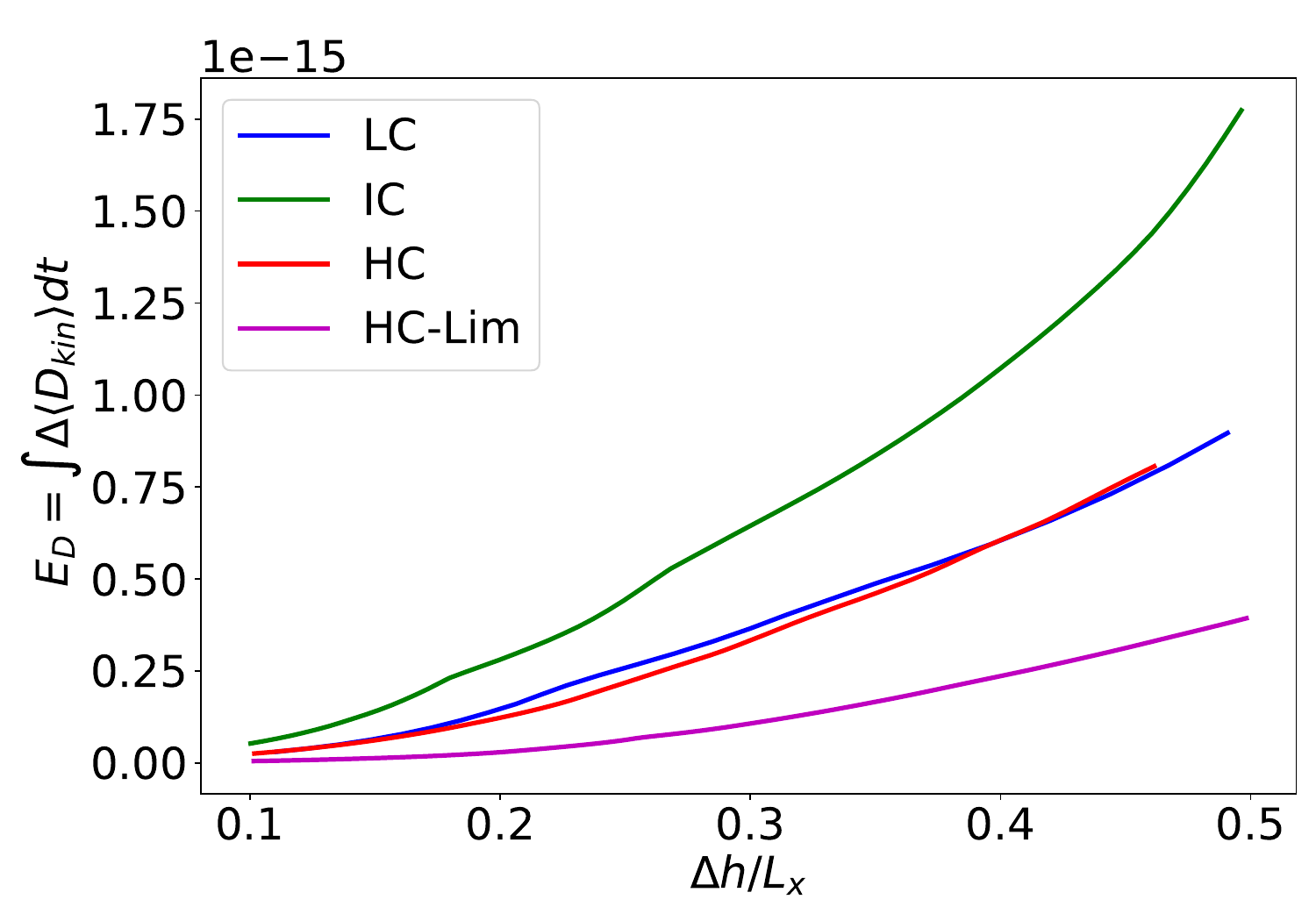}
\caption{Cumulative ion--neutral drag dissipation $E_{\rm kin}(t)=\int_0^t\langle D_{\rm kin}\rangle\,dt$.}
\end{subfigure}
\caption{Global energy pathways as functions of the normalized mixing layer thickness $\Delta h/L_x$ for different ion--neutral coupling strengths. The intermediate--coupling regime maximizes the conversion of gravitational energy into drag-related dissipation, while both weak and strong coupling lead to reduced dissipation efficiency.}
\label{fig:global_energy}
\end{figure}

To complement the local force maps, we now examine the global energy pathways that govern the nonlinear evolution of the mixing layer. Figure~\ref{fig:global_energy} shows the cumulative gravitational energy injection $E_g(t)=\int_0^t\langle P_g\rangle\,dt$, with $P_g=\rho_c v_{c,z} g$, together with the cumulative ion--neutral drag dissipation $E_{\rm kin}(t)=\int_0^t\langle D_{\rm kin}\rangle\,dt$, where $D_{\rm kin}=\alpha\,\rho_c\rho_n|\mathbf{v}_c-\mathbf{v}_n|^2$, both represented as functions of the normalized mixing layer thickness $\Delta h/L_x$. A robust result is that the intermediate--coupling (IC) regime stands out energetically: at a given nonlinear stage, IC simultaneously reaches a maximum of drag-related dissipation and a minimum of gravitational energy content. This behavior is physically expected, since the gravitational source term scales with the charged velocity, $P_g\propto v_{c,z}$, so that efficient momentum exchange with neutrals reduces the charged flow speed and limits the gravitational energy retained in the charged component, while maximizing the relative slip between the two fluids and therefore the drag dissipation. As a result, the gravitational energy injected into the system is converted most efficiently into ion--neutral drift in the IC regime. In contrast, the weakly coupled (LC) and strongly coupled (HC) regimes display comparable global energetic behavior, both characterized by lower dissipation efficiencies. In the LC case, the charged fluid evolves almost independently of the neutrals, limiting the development of significant inter-fluid drift, whereas in the HC limit the two fluids are nearly locked together, suppressing relative motion and reducing drag dissipation. The HC-limit case further exhibits the largest gravitational energy content, consistent with the persistence of higher charged velocities when slip is strongly inhibited. It is important to emphasize that these simulations are performed in an open vertical configuration, so that the total energy budget is not strictly closed and energy can enter or leave the domain through the boundaries. For this reason, the present analysis does not aim at establishing a strict conservation law, but rather at comparing how gravitational energy is redistributed between physical channels across coupling regimes. From this perspective, the global trends reinforce the interpretation suggested by the local force maps: the nonlinear organization of the mixing layer is controlled not by the total amount of injected gravitational energy, but by how efficiently this energy is redirected into ion--neutral drift versus magnetic stresses.

We deliberately limit the analysis to this stage of the nonlinear evolution and do not attempt to characterize a late-time saturated regime. In magnetized Rayleigh–Taylor configurations, saturation does not necessarily coincide with the onset of fully developed turbulence, as is often assumed in purely hydrodynamic studies, but may instead result from a geometry- or force-balance–driven arrest that depends on the wavelength content and on the horizontal magnetic component  $B_x$. Moreover, because the vertical boundaries act as open reservoirs, the global energy budget is not strictly closed, so that extending the energy-based analysis to later times would require a dedicated control-volume approach tied to the mixing layer thickness or fully three-dimensional simulations with carefully designed boundary conditions. For these reasons, we focus here on the physically robust regime where the nonlinear organization of the mixing layer can be unambiguously related to the partition of injected gravitational power between ion–neutral drift and magnetic stresses.

\section{Conclusion}\label{sec:Conclusion}

In this work, we have investigated the nonlinear development of the Rayleigh--Taylor instability in a partially ionized plasma, with a particular emphasis on the role of ambipolar diffusion in the presence of an oblique magnetic field. Building on the linear analysis presented in \citetalias{Callies2025}, we combined high--resolution two--fluid simulations with self--consistent diagnostics in order to quantify how ion--neutral coupling modifies both the growth and the morphology of the mixing layer beyond the linear regime. We first demonstrated that our numerical setup accurately reproduces the predictions of the linear theory across hydrodynamic, single--fluid MHD, and bi--fluid configurations. Using a finger--bubble diagnostic, we confirmed that the theoretical growth rates are recovered for a wide range of wavelengths and coupling strengths, and that compressibility effects remain negligible in the parameter regime considered. This validation step provides a robust foundation for the nonlinear analysis. In the nonlinear regime, we showed that while the classical quadratic scaling of the mixing height, $h(t) \propto A g t^{2}$, remains a useful global description, it fails to capture important time--dependent features induced by ambipolar diffusion. By introducing a morphology--based alignment procedure based on a fixed normalized mixing height \( h/\lambda \), we were able to perform meaningful comparisons between simulations with vastly different linear growth rates. This approach revealed a characteristic inflected behavior of the mixing--layer evolution when ambipolar diffusion becomes dynamically important. In particular, we identified a specific coupling regime in which ambipolar diffusion produces a non--trivial modification of the nonlinear dynamics. After synchronization at an equivalent morphological stage, the evolution exhibits a transient phase of enhanced growth relative to the uncoupled reference, followed by a pronounced slowdown that leads to a reduced mixing efficiency at later times. The ratio of mixing heights between coupled and uncoupled runs asymptotically approaches a value significantly below unity, of order \( \sim 0.7 \) for the most affected cases. This behavior demonstrates that ambipolar diffusion does not simply rescale the nonlinear growth coefficient, but instead reshapes the temporal evolution of the instability in a genuinely time--dependent manner.

We further showed that these effects persist when the interface is perturbed over a broad range of wavelengths. In the multi--mode regime, ambipolar diffusion strongly modifies the morphology of the mixing layer, inhibiting large--scale plume coalescence and promoting a more fragmented interface at intermediate coupling. Although one--dimensional power spectra of the charge density at fixed nonlinear stage do not reveal a clear change in global scaling laws, they nonetheless confirm that the statistical distribution of power across horizontal wavenumbers remains largely insensitive to coupling. This absence of a strong spectral signature indicates that the observed morphological differences are not associated with a simple spectral cutoff, but rather reflect a local and geometrical reorganization of the flow. A more direct physical interpretation emerges from the analysis of energy conversion channels. By comparing the relative contributions of gravitational injection, Lorentz work, and ion--neutral drag dissipation at equivalent nonlinear stages, we showed that ambipolar diffusion primarily acts by redistributing energy locally within the mixing layer. In particular, the fraction of gravitational power converted into ion--neutral drift reaches a maximum at intermediate coupling, while the global contribution of magnetic stresses remains comparatively modest and weakly dependent on the coupling strength. This non--monotonic behavior provides a natural explanation for the enhanced smoothness and coherence observed in the intermediate regime, where the drift efficiently extracts energy from the flow without completely suppressing the development of large--scale structures.

Taken together, these results demonstrate that ambipolar diffusion introduces a bounded regime of nonlinear behavior in which the Rayleigh--Taylor instability departs qualitatively from the standard self--similar picture. Rather than acting as a simple effective diffusivity, ambipolar coupling produces an early acceleration followed by a sustained nonlinear braking, accompanied by a redistribution of energy between bulk motions, magnetic stresses, and differential ion--neutral drift. The resulting morphology reflects a delicate balance between magnetic tension, which suppresses short--wavelength perturbations, and ambipolar diffusion, which locally reorganizes the flow through slip--mediated dissipation. From a methodological standpoint, this work highlights the importance of morphology--based synchronization when comparing nonlinear evolutions across different physical regimes. Aligning simulations at a fixed normalized mixing height provides a robust framework to disentangle genuine physical effects from trivial timing differences associated with distinct linear growth rates. This approach is particularly well suited to multi--fluid systems, where additional timescales associated with coupling can strongly affect the apparent nonlinear evolution. Finally, while the present study focuses on idealized two--dimensional configurations, the mechanisms identified here are expected to play an important role in a wide range of astrophysical environments where partially ionized plasmas and magnetic fields coexist, such as molecular clouds, supernova remnants, or the interfaces of expanding H\,\textsc{ii} regions. Extensions to three dimensions, as well as to regimes including additional physical ingredients such as cooling, stratification, or cosmic--ray feedback, constitute natural directions for future work. The results presented here provide a physically grounded reference for such studies and emphasize that ambipolar diffusion can qualitatively alter the nonlinear outcome of classical hydromagnetic instabilities.

\begin{acknowledgements}
The authors are grateful to K. Ferrière, P. Lesaffre, F. Boulanger, O. Berné for fruitful discussions. We also thank A. Cogez for his contribution during his internship. This work was supported by the Thematic Action “Programme National Physique Stellaire” (PNPS) of INSU Programme National “Astro”, with contributions from CNRS Physique \& CNRS Chimie, CEA, and CNES. 
\end{acknowledgements}

\appendix 

\section{Discussion on \citetalias{Callies2025}}
\label{app:paper1_corrigendum}
In \citetalias{Callies2025}, the analytical dispersion relation was derived under the simplifying assumption that the ion--neutral and neutral--ion collision frequencies were identical on both sides of the density interface. While this assumption facilitates the algebraic treatment of the matching conditions, it is not fully consistent with the numerical prescription adopted in MPI--AMRVAC, where the drag coefficient $\alpha$ is conserved and the collision frequencies scale with the local density. In the numerical model, the collision frequencies are defined as
\begin{equation}
\nu_{\rm nc} = \alpha \rho_{\rm c},
\qquad
\nu_{\rm cn} = \alpha \rho_{\rm n},
\end{equation}
so that they vary proportionally to the density of the interacting species. Across an interface separating two homogeneous layers of densities $\rho_1$ and $\rho_2$, we define the density contrast
\begin{equation}
d \equiv \frac{\rho_1}{\rho_2}.
\end{equation}
Because the drag coefficient $\alpha$ is constant, the collision frequencies on either side of the interface scale in the same way, yielding
\begin{equation}
\nu_{\rm nc,2} = d \, \nu_{\rm nc,1},
\qquad
\nu_{\rm cn,2} = d \, \nu_{\rm cn,1}.
\end{equation}

To account for this asymmetry in the analytical formulation, we introduce the parameter
\begin{equation}
\xi = \frac{\nu_{\rm nc,1}}{\nu_{\rm nc,2}}
     = \frac{\nu_{\rm cn,1}}{\nu_{\rm cn,2}}.
\end{equation}
In the physically consistent case corresponding to a constant drag coefficient, this parameter is therefore not free but directly related to the density contrast as
\begin{equation}
\xi = d.
\end{equation}

The case $\xi = 1$ corresponds to the simplifying assumption adopted in \citetalias{Callies2025}, whereas $\xi = d$ reflects the collision-frequency variation implied by the numerical prescription. In the following, we reformulate the matching conditions accordingly.

:
\begin{align}
 M= \begin{pmatrix}
        1&1&0&-1&-1&0&0&0\\
        ik_x/k&0&1&ik_x/k&0&-1&0&0\\
        k&m_{1-}&0&k&-m_{2+}&0&0&0\\
        k^2&m_{1-}^2&0&-k^2&-m_{2+}^2&0&0&0\\
        ik_x&0&m_{1-}&-ik_x&0&-m_{2+}&0&0\\
        \alpha_{1{\rm c}}&\beta_{1{\rm c}}&\gamma_{1{\rm c}}&-\alpha_{2{\rm c}}&-\beta_{2{\rm c}}&-\gamma_{2{\rm c}}& \delta_{1{\rm c}}&-\delta_{2{\rm c}}\\
        0&\psiOne&0&0&-\psiTwo&0&1&-1\\
       \alpha_{1{\rm n}}&\beta_{1{\rm n}}&0&-\alpha_{2{\rm n}}&-\beta_{2{\rm n}}&0& \delta_{1{\rm n}}&-\delta_{2{\rm n}}\\
    \end{pmatrix} \ ,
\end{align}

with
\begin{align}
     \phiOne &=\sqrt{\dfrac{\Omega+\nucn}{\Omega +\nucn+\nunc}}\\
    \phiTwo &=\sqrt{\dfrac{\xi\Omega+\nucn}{\xi\Omega +\nucn+\nunc}}\\
    \psiOne &=\dfrac{\nucn}{\nucn+\Omega}\\
    \psiTwo &=\dfrac{\nucn}{\nucn+\xi\Omega}\\
    m_{1-}&=-\left[i\left(\dfrac{k_x}{\tan\theta}\right)-\dfrac{\Omega}{\sin\theta\,\cac\phiOne}\right]\\
    m_{2+}&=-\left[i\left(\dfrac{k_x}{\tan\theta}\right)+\dfrac{\Omega}{\sin\theta\,\sqrt{d}\cac\phiTwo}\right]
\end{align}
\begin{align}
    \alpha_{1{\rm c}}&\equiv k_x\left[\gc-\left(\cAone^2k\sin^2\theta-\dfrac{\Omega\left(\nunc+\Omega\right)}{k}\right)\right] \ ,\\
    \alpha_{2{\rm c}}&\equiv d^{-1}k_x\left[ \gc+\left(d\cAone^2k\sin^2\theta-\dfrac{\Omega\left(\nunc/\xi+\Omega\right)}{k}\right)\right] \ ,\\
    \beta_{1{\rm c}}&\equiv k_x\gc \ ,\qquad \beta_{2{\rm c}}\equiv d^{-1}k_x\gc \ ,\\
    \gamma_{1{\rm c}}&\equiv i\left(\cAone^2m_{1-}^2\sin^2\theta-\phiOne^{-2}\Omega^2\right) \ ,\\
    \gamma_{2{\rm c}}&\equiv d^{-1}i\left(d\cAone^2m_{2+}^2\sin^2\theta-\phiTwo^{-2}\Omega^2\right) \ ,\\
    \delta_{1{\rm c}}&\equiv -k_x\dfrac{\Omega\nunc}{k}\ , \qquad \delta_{2{\rm c}}\equiv d^{-1}k_x\dfrac{\Omega\nunc/\xi}{k} \ ,
\end{align}
\begin{align}
    \alpha_{1{\rm n}}&\equiv -k_x\dfrac{\Omega\nucn}{k} \ ,\\
    \alpha_{2{\rm n}}&\equiv d^{-1}k_x\dfrac{\Omega\nucn/\xi}{k} \ ,\\
    \beta_{1{\rm n}}&\equiv k_x\psiOne \gn \ , \qquad \beta_{2{\rm n}}\equiv d^{-1}k_x\psiTwo \gn \ ,\\
    \delta_{1{\rm n}}&\equiv k_x\left(\gn+\dfrac{\Omega(\nucn+\Omega)}{k}\right) \ ,\\
    \delta_{2{\rm n}}&\equiv d^{-1}k_x\left(\gn-\dfrac{\Omega(\nucn/\xi+\Omega)}{k}\right) \ .
\end{align}

Allowing the collision frequencies to vary across the interface modifies the analytical formulation in a controlled but non--trivial way. When $\xi\neq1$, the neutral--ion and ion--neutral collision frequencies explicitly differ on each side of the interface, so that the matching conditions involve distinct dynamical responses in the two layers. As a result, several auxiliary quantities introduced in \citetalias{Callies2025}, such as the parameters $\phi_{1,2}$ and $\psi_{1,2}$, acquire an explicit dependence on $\xi$, and the coefficients entering the interface conditions---in particular $\alpha_{i{\rm c}}$, $\beta_{i{\rm c}}$, $\gamma_{i{\rm c}}$, $\delta_{i{\rm c}}$ and their neutral counterparts---must be evaluated using different collision frequencies on either side of the interface. While the overall block structure of the matching matrix is preserved, this asymmetry breaks part of the algebraic simplifications exploited in \citetalias{Callies2025}, where the coupling could be represented by a single effective parameter. In the present formulation, the coupling enters through several coefficients simultaneously, which makes the dispersion relation less amenable to a compact analytical reduction and renders the interpretation of the coupling regimes less direct from the mathematical expression alone. Nevertheless, the physical structure of the problem remains unchanged: the dispersion relation still connects continuously the limits of vanishing coupling and perfect coupling, and the same asymptotic behaviours are recovered when the collision frequency becomes either negligible or dominant compared to the dynamical timescale.

\begin{figure}
    \centering
    \includegraphics[width=\linewidth]{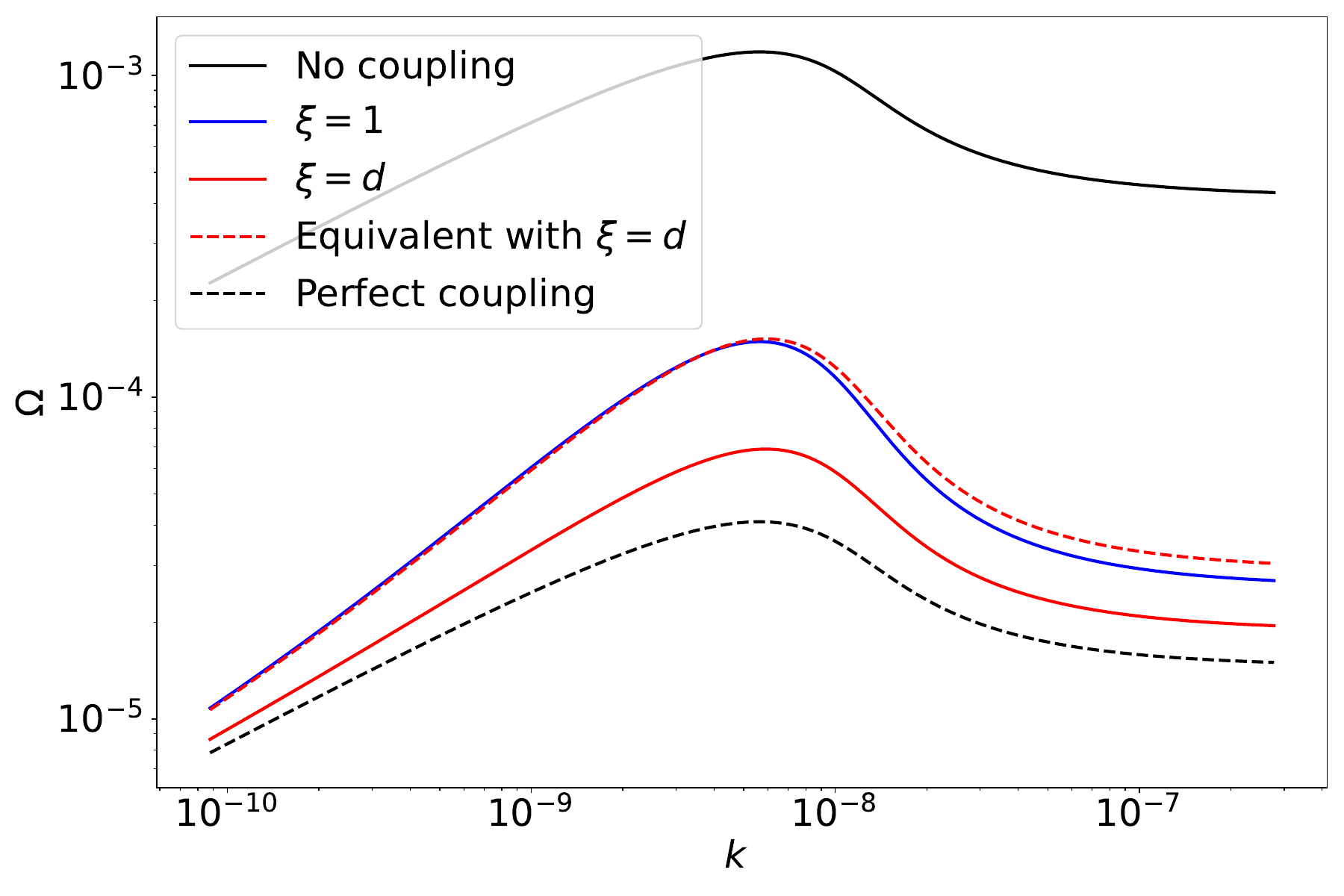}
\caption{Linear growth rate as a function of the wavenumber $k$.
The black curves show the two asymptotic limits: the uncoupled case (``No coupling'') and the perfectly coupled case (``Perfect coupling''). The blue curve corresponds to the formulation with $\xi=1$ used in \citetalias{Callies2025}, while the red curve shows the physically consistent case $\xi=d$, where the collision frequencies scale with the density contrast across the interface (constant drag coefficient $\alpha$).
The dashed red curve labelled ``Equivalent with $\xi=d$'' shows the $\xi=1$ formulation evaluated with an effective collision frequency $\nu_{\rm nc,eq}$ chosen so as to reproduce the $\xi=d$ dispersion curve over the intermediate-$k$ range.}
    \label{fig:corrigendum}
\end{figure}

The practical impact of this modification is illustrated in Fig.~\ref{fig:corrigendum}, where the linear growth rate is plotted as a function of the wavenumber $k$ for different coupling strengths, comparing the cases $\xi=1$ (as assumed in \citetalias{Callies2025}) and $\xi=d$, corresponding to a collision frequency variation driven solely by the density contrast across the interface. The uncoupled and fully coupled asymptotic limits are shown for reference and are identical in both cases, confirming that the modification does not alter the fundamental limiting behaviour of the instability. Differences arise primarily at intermediate wavenumbers, where the growth rate obtained with $\xi=d$ is systematically reduced compared to the $\xi=1$ case for the same nominal collision frequency. This reduction reflects the asymmetric collision-frequency distribution across the interface: when $\xi=d$, the coupling strength differs between the two layers, which modifies the relative contribution of each fluid to the linear dynamics and enhances the effective damping at intermediate scales. For practical comparison with the formulation of \citetalias{Callies2025}, it is convenient to introduce an effective collision frequency $\nu_{\rm nc,eq}$ defined operationally as the value that minimises the mismatch between the $\xi=d$ dispersion curve and the $\xi=1$ curve over a chosen intermediate-$k$ interval. With this definition, the two formulations can be brought into close quantitative agreement without affecting the uncoupled and perfectly coupled limits. The ratio $\nu_{\rm nc,eq}/\nu_{\rm nc}$ is not universal: it depends on the density contrast $d$ and, to a lesser extent, on the coupling regime considered. Therefore, the transition from $\xi=1$ to $\xi=d$ does not introduce a new qualitative regime of instability, but rather corresponds to a renormalisation of the effective coupling strength when the collision frequencies are allowed to scale consistently with density. The theoretical framework developed in \citetalias{Callies2025} thus remains fully applicable, provided that collision frequencies are interpreted in terms of an effective coupling when comparing analytical predictions with physically consistent numerical prescriptions.

\end{document}